\documentclass[twocolumn,pra,superscriptaddress,floatfix]{revtex4-2}
\pdfoutput=1 
\usepackage{amsmath}
\usepackage{amsfonts}
\usepackage{amssymb}
\usepackage{physics}
\usepackage{graphicx}
\usepackage{bbm}
\usepackage[dvipsnames]{xcolor}
\allowdisplaybreaks

\begin{document}

\title{Ghost Factors in Gauss Sum Factorization with Transmon Qubits}

\author{Lin Htoo \surname{Zaw}}
\altaffiliation[Present address: ]{Centre for Quantum Technologies, National University of Singapore, 3 Science Drive 2, Singapore 117543, Singapore}
\affiliation{Division of Physics and Applied Physics, School of Physical and Mathematical Sciences, Nanyang Technological University, 21 Nanyang Link, Singapore 637371, Singapore}

\author{Yuanzheng Paul \surname{Tan}}
\affiliation{Division of Physics and Applied Physics, School of Physical and Mathematical Sciences, Nanyang Technological University, 21 Nanyang Link, Singapore 637371, Singapore}

\author{Long Hoang \surname{Nguyen}}
\affiliation{Division of Physics and Applied Physics, School of Physical and Mathematical Sciences, Nanyang Technological University, 21 Nanyang Link, Singapore 637371, Singapore}

\author{Rangga P. \surname{Budoyo}}
\affiliation{Centre for Quantum Technologies, National University of Singapore, 3 Science Drive 2, Singapore 117543, Singapore}

\author{Kun Hee \surname{Park}}
\affiliation{Centre for Quantum Technologies, National University of Singapore, 3 Science Drive 2, Singapore 117543, Singapore}

\author{Zhi Yang \surname{Koh}}
\affiliation{Division of Physics and Applied Physics, School of Physical and Mathematical Sciences, Nanyang Technological University, 21 Nanyang Link, Singapore 637371, Singapore}

\author{Alessandro \surname{Landra}}
\altaffiliation[Present address: ]{IQM Finland Oy, Keilaranta 19, 02150 Espoo, Finland}
\affiliation{Centre for Quantum Technologies, National University of Singapore, 3 Science Drive 2, Singapore 117543, Singapore}

\author{Christoph \surname{Hufnagel}}
\affiliation{Centre for Quantum Technologies, National University of Singapore, 3 Science Drive 2, Singapore 117543, Singapore}

\author{Yung Szen \surname{Yap}}
\affiliation{Centre for Quantum Technologies, National University of Singapore, 3 Science Drive 2, Singapore 117543, Singapore}
\affiliation{Faculty of Science and Centre for Sustainable Nanomaterials (CSNano), Universiti Teknologi Malaysia, 81310 UTM Johor Bahru, Johor, Malaysia}

\author{Teck Seng \surname{Koh}}
\thanks{Corresponding author:~\href{mailto:kohteckseng@ntu.edu.sg}{kohteckseng@ntu.edu.sg}}
\affiliation{Division of Physics and Applied Physics, School of Physical and Mathematical Sciences, Nanyang Technological University, 21 Nanyang Link, Singapore 637371, Singapore}
\email{kohteckseng@ntu.edu.sg}

\author{Rainer \surname{Dumke}}
\thanks{Corresponding author:~\href{mailto:rdumke@ntu.edu.sg}{rdumke@ntu.edu.sg}}
\affiliation{Division of Physics and Applied Physics, School of Physical and Mathematical Sciences, Nanyang Technological University, 21 Nanyang Link, Singapore 637371, Singapore}
\affiliation{Centre for Quantum Technologies, National University of Singapore, 3 Science Drive 2, Singapore 117543, Singapore}
\email{rdumke@ntu.edu.sg}

\begin{abstract}
A challenge in the Gauss sums factorization scheme is the presence of ghost factors --- nonfactors that behave similarly to actual factors of an integer --- which might lead to the misidentification of nonfactors as factors or vice versa, especially in the presence of noise. We investigate Type II ghost factors, which are the class of ghost factors that cannot be suppressed with techniques previously laid out in the literature. The presence of Type II ghost factors and the coherence time of the qubit set an upper limit for the total experiment time, and hence the largest factorizable number with this scheme. Discernability is a figure of merit introduced to characterize this behavior. We introduce preprocessing as a strategy to increase the discernability of a system, and demonstrate the technique with a transmon qubit. This can bring the total experiment time of the system closer to its decoherence limit, and increase the largest factorizable number.
\end{abstract}

\maketitle
\section{Introduction}
Qubit interference plays an important role in quantum algorithms, which can be used to solve some problems more efficiently than classical algorithms \cite{Cleve1998,Bennett2000,Lloyd2000,Stahlke2014}. However, on present Noisy Intermediate-Scale Quantum (NISQ) era systems, noisy qubits are and will be prevalent, so arbitrarily long operations might be impractical on currently available architectures \cite{Preskill2018}. As such, quantum systems might lose their advantage over their classical counterparts due to decoherence effects.

A prime example on the impact of decoherence on quantum systems is Shor's algorithm, which relies on the interference of superposed qubits in an intermediate step \cite{Shor1999}. What on paper is a scheme that can outperform classical algorithms requires more than $2\log_2(N)$ qubits for the factorization of an integer $N$ \cite{Beauregard2003, Haner2017}. This can be challenging due to the coherence times of present qubit technology, limiting experimental implementations to two-digit numbers \cite{Skosana2021,MartinLopez2012}. In fact, for a fault-tolerant execution of Shor's algorithm with noisy qubits, the number of physical qubits needed greatly exceeds the ideal value required \cite{Devitt2013}.

Therefore, textbook quantum schemes have to be tailored for use with NISQ-era systems in order to boost their applicability even in the face of noisy operations. In this paper, we adapt an integer factorization technique, one that utilizes the destructive interference of quantum phases, for use in a transmon qubit, and we limit the scope of this work to a single qubit.

Apart from Shor's algorithm, other approaches to integer factorization have been suggested and demonstrated, including the use of quantum annealing \cite{Peng2008, Dridi2017, Jiang2018}, variational algorithms \cite{Anschuetz2018, Qiu2020}, and Gauss sums \cite{Mack2002,Merkel2006,Rangelov2009}. In this paper, we turn to the lattermost technique for two reasons. Firstly, Gauss sum factorization have been demonstrated on many architectures, including nuclear magnetic resonance (NMR) systems \cite{Mahesh2007,Mehring2007}, Bose-Einstein condensates \cite{Sadgrove2008}, multipath interferometers \cite{Tamma2011}, with the optical Talbot effect \cite{Bigourd2008, Pelka2018}, and a variation has been proposed for Josephson phase qubits \cite{Ng2010}. Its widespread use means that our findings should be applicable to a variety of systems. Secondly, while it does not provide a speedup over the classical case on its own \cite{Merkel2011}, there are extended Gauss sum schemes with the promise of bringing together the robustness of Gauss sums with the speedup of Shor's algorithm \cite{Jun2011,Wolk2012}. As coherent control of entanglement between a larger number of qubits become possible, although not yet in the scale required for the direct implementation of textbook schemes, these hybrid schemes can be immediately useful.

The naive form of Gauss sum factorization takes advantage of the interference behavior of a quantum system, where the number to be factorized and various trial factors are worked into the parameters that govern the system's evolution \cite{Wolk2011}. The protocol is designed so that the system constructively interferes when the trial factor is a factor, but destructively interferes if the trial factor is a nonfactor.

For Gauss sum factorization to be effective in NISQ-era architectures, the challenge of ghost factors have to be addressed. Ghost factors are nonfactors that behave similarly to actual factors, whose presence might cause the misidentification of factors as nonfactors or vice versa. Their scaling behavior have been previously studied, and it is known that to subdue the effect of ghost factors, the ``truncation parameter'' --- the number of unitary gates needed --- has to be above a lower limit that scales with $N$ \cite{Stefanak2007}. Monte Carlo methods to suppress ghost factors have also been utilised \cite{Peng2008,Weber2008}.

However, as noted by \citet{Stefanak2007}, these methods only suppress a class of ghost factors, which we call \emph{Type I} ghost factors. There is a gap in the examination of what we designate as \emph{Type II} ghost factors, which are ghost factors not suppressed with the aforementioned methods. These Type II ghost factors become prevalent when the total experimental time is close to the coherence limit of the qubit. As the truncation parameter must be large for a large $N$, Type II ghost factors greatly restrict the size of the number that can be factorized. As such, it is of importance to also subdue Type II ghost factors to push the utility of the quantum schemes, even in the present NISQ era of noisy qubits and imperfect operations.

In this paper, we investigate the impact of Type II ghost factors on the Gauss sum factorization scheme, supplemented by both theoretical arguments and experimental demonstrations. The experiments are performed on a transmon qubit, and our experimental setup is detailed in Sec.~\ref{sec:experimental-setup}. In Sec.~\ref{sec:theory}, we lay out the theory of Gauss sums factorization, the Bloch-Redfield model to describe qubit decoherence, and the appearance of ghost factors. We show that Type II ghost factors limit the effectiveness of the scheme even where Type I ghost factors are suppressed, and the decoherence when the time of the experiment approaches the coherence time of the qubit sets an upper limit to the computation. We introduce a figure of merit that characterizes this upper limit, which we call \emph{discernability}, in Sec.~\ref{sec:discernability}.

In Sec.~\ref{sec:preprocessing}, we introduce and experimentally demonstrate preprocessing as a technique to improve the ability of a system to discern factors and nonfactors in a Gauss sum factorization experiment, to push the utility of the scheme even as the system approaches its decoherence limit. Finally, in Sec.~\ref{sec:fit-more-pulse}, we address a discrepancy we encountered in our experimental results which required a phenomenological adjustment to the noise model, and suggest a possible direction for future works.

\section{Experimental Setup\label{sec:experimental-setup}}

\begin{figure}
    \includegraphics[width=.4\textwidth]{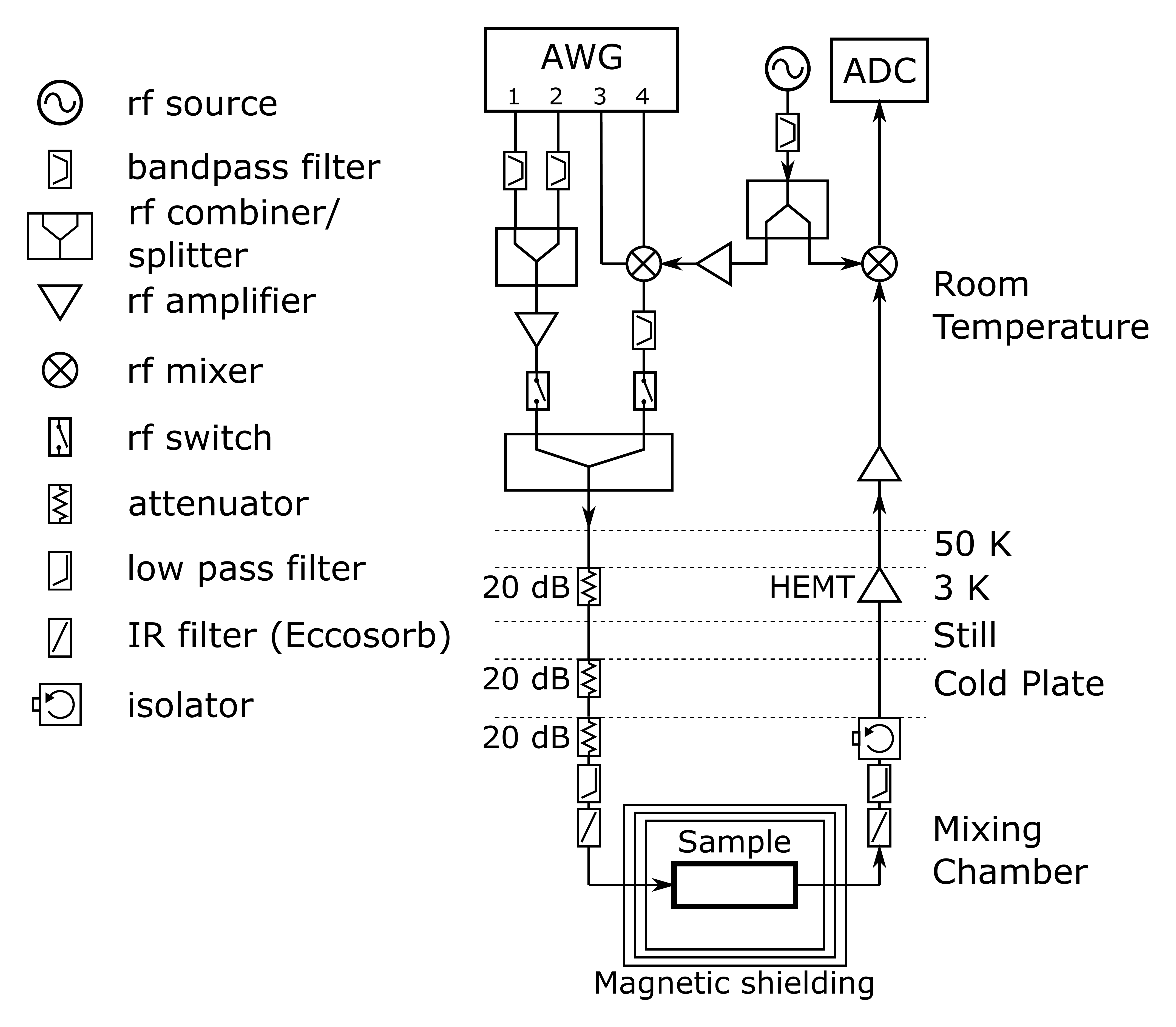}
    \caption{\label{fig:experimental-setup} Block diagram of the experimental setup. The figure is divided into two parts, the room temperature electronics at the top, and the cyrogenic electronics and the cryostat at the bottom. An arbitrary waveform generator (AWG) is used to synthesize the signals for qubit control and readout. The readout signals are upconverted and downconverted with an IQ mixer and a local oscillator. Both signals are combined and sent through the cryostat. The readout signal is then downconverted and parsed by an analog-to-digital converter (ADC).}
\end{figure}

In Fig.~\ref{fig:experimental-setup}, the schematic diagram of the experimental setup is shown. Two transmon qubits are located in a rectangular copper 3D cavity with a bare resonant frequency of  $4.517~\mathrm{GHz}$. The Q factor of the cavity is at 3050. The two transmons are thermalized at 9-12 mK and have resonant frequencies $\omega_q^{(A)} = 3.064~\mathrm{GHz}$ and $\omega_q^{(B)} = 3.266~\mathrm{GHz}$. All experiments were performed with qubit A. The $T_1$ time of qubit A is $4.7\pm0.5 ~\mu\text{s}$. The $T_2$ time of qubit A is $1.9\pm0.2 ~\mu\text{s}$. For qubit A, using state tomography, we determine the $\ket{0}$ state fidelity to be 99.3\% and the $\ket{1}$ state fidelity to be 98.9\%. An arbitrary waveform generator (AWG5204 by Tektronix) is used to synthesize the pulses for qubit control and readout. Fixed amplitude single qubit gate operations vary in duration from 9 to 20 ns. Each experimental pulse sequence consists of the pulse train, followed by a 5~$\mu$s readout pulse and a 20~$\mu$s rest period for the qubit and cavity reset. This readout length provides the best signal to noise ratio for readout.

\section{Theory\label{sec:theory}}

\subsection{Gauss Sum Factorization}
Factorization using Gauss sums is motivated by the observation that the term $\exp(i2\pi m^2 \frac{N}{l})$ is one for a sequence of $m=0,1,\dots$ when $N/l$ is an integer, but oscillates widely around the origin if not. Hence, for a number $N$ and a trial factor $l$, the average of the terms add up to unity when $l$ is a factor of $N$, but otherwise destructively interferes to a value less than one.

\begin{figure}
  \centering\includegraphics{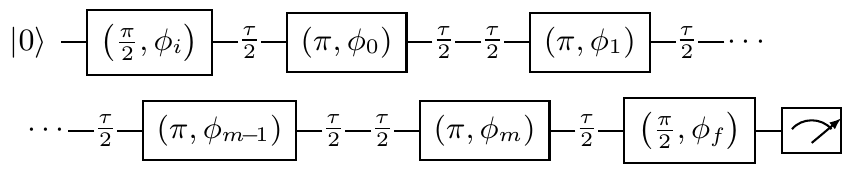}
  \caption{\label{fig:circuit}Pulse sequence used in Gauss sum factorization. Most operations are $\pi$ rotations around $\cos(\phi_k)\hat{x}+\sin(\phi_k)\hat{y}$, where $\phi_k$ parameterizes the axis of rotation. The initialization pulses are $\pi/2$ rotations about $\phi_i$ and $\phi_f$ respectively. In our experiment, $\phi_i = \phi_f = \pi/2$, and the measurement returns the probability of finding the system in the state $\ket{1}$. There is a wait time of $\tau$ between successive pulses.}
\end{figure}

Furthermore, recognizing that these terms are akin to the phase factors that accumulate when operating a qubit in the equatorial plane of the Bloch sphere, a particular choice of pulse sequences can generate states whose phases are the Gauss sum terms. We adopt the pulse sequence laid out by \citet{Mehring2007}, given as a sequence of $\pi$-pulses about the axis $\cos(\phi_k)\hat{x}+\sin(\phi_k)\hat{y}$ with
\begin{equation}\label{eq:gauss-angles}
  \begin{aligned}
    \phi_k(l) &= \begin{cases}
    0 & \text{for $k=0$}; \\
    a_k(N)/l & \text{for $k>0$},\\
  \end{cases}\\
  a_k(N) &= \pqty{-1}^k\pi N\pqty{2k-1}
  \end{aligned}
\end{equation}
where $N$ is the number to be factorized, and $l$ is the trial factor. This sequence, as shown in Fig.~\ref{fig:circuit}, gives
\begin{align}
  \Pr(m;l,N) &= \ev{\prod_{k=m}^0 \hat{R}(\pi,\phi_k)}{+} \nonumber\\
  &=\frac{1}{2}\pqty{ 1 + c^{(m)}_N(l) },\\
  \ev{\Pr(l,N)}^{(M)} &= \frac{1}{M+1}\sum_{m=0}^M\Pr(m;l,N)\nonumber\\
  &= \frac{1}{2}\pqty{ 1 + C^{(M)}_N(l) },
\end{align}
where $M$ is the maximum number of pulses used for each trial factor. The details are worked out in Appendix \ref{apd:resultant-rotations}. The averaged probability $\ev{\Pr(l,N)}^{(M)}$ acts as the \emph{signal} of each trial factor in a Gauss sum factorization, and the magnitude of the signal indicates whether or not the trial factor is a factor or a nonfactor. Here, $c^{(m)}_N(l)$ and $C^{(M)}_N(l)$ are the real Gauss summands and sums,
\begin{align}
  c^{(m)}_N(l) &= \cos(2\pi m^2\frac{N}{l})\\
  C^{(M)}_N(l) &= \frac{1}{M+1}\sum_{m=0}^{M}c^{(m)}_N(l).\label{eq:gauss-sum}
\end{align}
The sum $C^{(M)}_N(l) = 1$ if $l$ is a factor of $N$, while $C^{(M)}_N(l) < 1$ if $l$ is not a factor of $N$. By performing the sum with trial factors $1 \leq l \leq \sqrt{N}$ and seeking out the peaks, we can work out the factors of a large number.

However, in practice, the presence of noise makes it difficult to discern between factors and nonfactors, as fluctuations and decays would make the Gauss sum of a factor less than one. Furthermore, there are \emph{ghost factors} --- nonfactors with Gauss sums that are very close to one, which might cause a misidentification of a ghost factor as a factor or vice versa. Therefore, in the experimental implementation of Gauss sums, the effect of noise must be considered.

\vspace{5ex}

\subsection{Bloch-Redfield Theory}
To study the effect of decoherence on Gauss sums, we turn to the Bloch-Redfield master equation that describes the evolution of an initial state $\hat{\rho} = \ketbra{\psi}{\psi}$, where $\ket{\psi} = \alpha\ket{0} + \beta\ket{1}$, that is weakly coupled to a noisy environment \cite{Krantz2019}.
\begin{equation}
\hat{\rho}_{\text{BR}}
= \pmqty{
1 + \pqty{\abs{\alpha}^2 - 1}e^{-\Gamma_1t} &
\alpha\beta^*e^{i\delta\omega t}e^{-\Gamma_2 t} \\
\alpha^*\beta e^{-i\delta\omega t}e^{-\Gamma_2 t} &
\abs{\beta}^2 e^{-\Gamma_1 t}
}
\end{equation}
Here, $\delta\omega = \omega_q-\omega_d$ is the difference between the qubit and driving frequency, $\Gamma_{1}=1/T_{1}$ with the relaxation time $T_1$, and $\Gamma_{2} = 1/T_{2}$ with the dephasing time $T_2$.

A density operator $\hat{\rho}$ can also be vectorized into a column vector $\ket{\varrho}$ as
\begin{equation}
\ket{\varrho} = \operatorname{vec}(\hat{\rho}) \equiv \frac{1}{\sqrt{2}}\pmqty{
1 \\
\alpha\beta^* + \alpha^*\beta \\
i\pqty{\alpha\beta^* - \alpha^*\beta} \\
\abs{\alpha}^2 - \abs{\beta}^2
}
\equiv\frac{1}{\sqrt{2}}\pmqty{
1 \\
\vec{r}
}
\end{equation}
where $\vec{r} = \pmqty{\ev{\hat{\sigma}_x},\ev{\hat{\sigma}_y},\ev{\hat{\sigma}_z}}^T$ with $0 \leq \abs{\vec{r}}^2 \leq 1$.

In this vectorized form, a unitary operator $\hat{R}(\theta,\hat{n})$ acting on a density operator $\hat{R}\hat{\rho}\hat{R}^\dagger$ is simply a matrix multiplication $\mathcal{R}(\theta,\hat{n})\ket{\varrho} = \spmqty{1&0\\0&R(\theta,\hat{n})}\ket{\varrho}$, where $R(\theta,\hat{n})$ is a 3D rotation matrix that acts on the Bloch vector $\vec{r}$.

Performing the same treatment on the Bloch-Redfield density operator,
\begin{widetext}
\begin{align}
\ket{\varrho_{\text{RB}}}
&=
\frac{1}{\sqrt{2}}
\pmqty{
	1\\
	\left( \alpha\beta^* e^{i\delta\omega t}+\alpha^*\beta e^{-i\delta\omega t}\right)e^{-\Gamma_2t} \\
	i\left( \alpha\beta^* e^{i\delta\omega t}-\alpha^*\beta e^{-i\delta\omega t}\right)e^{-\Gamma_2t} \\
	\left(1-e^{-\Gamma_1t}\right)+\left( {|\alpha|}^2-{|\beta|}^2 \right)e^{-\Gamma_1t}
}
=
\pmqty{
    1&0&0&0\\
    0&\cos(\delta\omega t)e^{-\Gamma_2 t}&\sin(\delta\omega t)e^{-\Gamma_2 t}&0\\
    0&-\sin(\delta\omega t)e^{-\Gamma_2 t}&\cos(\delta\omega t)e^{-\Gamma_2 t}&0\\
    1-e^{-\Gamma_1t}&0&0&e^{-\Gamma_1t}
}\ket{\varrho}
\nonumber\\[2ex]
&=
\underbrace{\pmqty{
    1&0&0&0\\
    0&\sqrt{e^{-\Gamma_1t}}&0&0\\
    0&0&\sqrt{e^{-\Gamma_1t}}&0\\
    1-e^{-\Gamma_1t}&0&0&e^{-\Gamma_1t}
}}_{{\text{amplitude damping with factor }e^{-\Gamma_1t}}}
\underbrace{\pmqty{
    1&0&0&0\\
    0&e^{-\pqty{\Gamma_2-\frac{\Gamma_1}{2}} t}&0&0\\
    0&0&e^{-\pqty{\Gamma_2-\frac{\Gamma_1}{2}} t}&0\\
    0&0&0&1
}}_{{\text{pure dephasing with factor }e^{-(\Gamma_2-\frac{1}{2}\Gamma_1) t}}}
\underbrace{\pmqty{
    1&0&0&0\\
    0&\cos(\delta\omega t)&\sin(\delta\omega t)&0\\
    0&-\sin(\delta\omega t)&\cos(\delta\omega t)&0\\
    0&0&0&1
}}_{z\text{-rotation of angle }\delta\omega t}\ket{\varrho}\\[2ex]
&\equiv
\mathcal{A}\pqty\big{e^{-\Gamma_1t}}
\mathcal{P}\pqty\big{e^{-(\Gamma_2-\frac{1}{2}\Gamma_1) t}}
\mathcal{R}\pqty\big{\delta\omega t,\hat{z}}
\ket{\varrho}.\nonumber
\end{align}
\end{widetext}
The operation $\mathcal{R}(\pi,\phi)$, a $\pi$-rotation around the axis $\cos(\phi)\hat{x} + \sin(\phi)\hat{y}$, commutes with both the amplitude damping and pure dephasing processes, while its commutation relation with a $z$-rotation is $\mathcal{R}\!\left(\delta\omega t,\hat{z}\right)\mathcal{R}(\pi,\phi) = \mathcal{R}(\pi,\phi)\mathcal{R}\!\left(-\delta\omega t,\hat{z}\right)$. So,
\begin{equation}\label{eq:bloch-redfield-decoherence}
\begin{aligned}
&\phantom{{}={}}\cdots
\mathcal{R}(\pi,\phi_{k-1})\mathcal{A}\mathcal{P}\mathcal{R}\!\left(\delta\omega t,\hat{z}\right)
\mathcal{R}(\pi,\phi_k)
\mathcal{A}\mathcal{P}\mathcal{R}\!\left(\delta\omega t,\hat{z}\right)
\cdots\\
&=
\cdots
\mathcal{R}(\pi,\phi_{k-1})\mathcal{R}(\pi,\phi_k)
\mathcal{A}^2\mathcal{P}^2\underbrace{\mathcal{R}\!\left(-\delta\omega t,\hat{z}\right)\mathcal{R}\!\left(\delta\omega t,\hat{z}\right)}_{\mathbbm{1}}
\cdots \\
&=
\cdots
\mathcal{R}(\pi,\phi_{k-1})\mathcal{R}(\pi,\phi_k)
\mathcal{A}^2\mathcal{P}^2
\cdots
\end{aligned}
\end{equation}
The expected measurement under this decay model is
\begin{equation}
\bra{\varrho_{+\hat{x}}}
\prod_{k=m}^0\mathcal{R}(\pi,\phi_k)\mathcal{A}^{m+1}\mathcal{P}^{m+1}\ket{\varrho_{+\hat{x}}}
\end{equation}
where $\ket{\varrho_{+\hat{x}}} = (1,1,0,0)^T/\sqrt{2}$ is the state in the $+\hat{x}$ axis, which comes from our choice of the initialization pulses. If the delay between pulses is $\tau$ and the $\pi$-pulse time is $t_\pi$, this gives
\begin{equation}
\widetilde{\Pr}(m;l,N) =\frac{1}{2}\pqty{
1 + \cos\pqty{2\pi m^2\frac{N}{l}}e^{-(m+1)\Gamma_2(\tau+t_\pi)}
}
\end{equation}
with the tilde indicating the presence of decoherence due to noise in the Bloch-Redfield model. Hence, the \emph{noisy} Gauss sum result is
\begin{equation}\label{eq:noisy-gauss-sum-br}
\widetilde{C}^{(M)}_N(l) = \frac{1}{M+1}\sum_{m=0}^M \cos\left(2\pi m^2\frac{N}{l}\right)e^{-(m+1)\Gamma_2(\tau+t_\pi)}.
\end{equation}
This is exactly the phenomenological fit used in an earlier Gauss sum experiment \cite{Mehring2007}.

\subsection{Ghost Factors}\label{sec:ghost-factors}
Eq.~\eqref{eq:gauss-sum} with $M=l-1$ includes every unique summand, since subsequent summands $m=k$ to $kl-1$ for $k=1,2,\dots$ are repetitions of the first sum. $M$ also corresponds to the maximum number of pulses used for each trial factor. Since the factorization of large $N$ requires large trial factors, it can be impractical to perform a complete Gauss sum due to decoherence limits. In practice, the \emph{truncated} Gauss sum with some constant $M < l_{\text{max}}-1$ is used.

The choice of $M$ is important as choosing a low number of pulses might not allow the summands to interfere enough to reduce the signal of a nonfactor adequately. This is especially true for larger trial factors, where many terms might be needed before it converges to a value close to its full sum. These appear as \emph{ghost factors}: nonfactors whose signals are close to that of a factor, which makes it difficult to differentiate the two. Previous theoretical work on truncated Gauss sums set a lower limit for the number of pulses required as $\sqrt[4]{N/4} \leq M$ \cite{Stefanak2007}.

As it turns out, this only addresses a subset of ghost factors. Here, it is useful to categorize the types of trial factors as
\begin{enumerate}
 \item \emph{Factors} that divide $N$, which includes both prime factors of $N$ and their products,
 \item \emph{``Well-behaved'' nonfactors} that decays quickly within a few pulses,
 \item \emph{Type I ghost factors} that take a large number of pulses to decay,
 \item \emph{Type II ghost factors} that plateau at some value.
\end{enumerate}
While Type I ghost factors can be suppressed by increasing the number of pulses $M$ used, the same cannot be done for Type II ghost factors, even with a large $M$. As such, we consider the effects of Type II ghost factors in the cases where Type I ghost factors are already suppressed (that is, the lower limit for $M$ is already met).

Any rational number can be reduce to the form $\frac{N}{l} = \text{integer} + \frac{p}{q}$, where $p < q$, $p$ and $q$ are coprime. Explicitly, $q = l/\gcd(N,l)$ and $p=(N/\gcd(N,l))\bmod{q}$. Then,
\begin{equation}\label{eq:gauss-sum-coprime}
C^{(M)}_N(l) = C^{(M)}_p(q) = \frac{1}{M+1}\sum_{m=0}^M \cos\left(2\pi m^2\frac{p}{q}\right),
\end{equation}
so $p$ and $q$ completely determine its behavior. Treating the cosine as the real part of a point in the complex plane, it is clear that the points traversed by this sum are the $q$-th roots of unity.

Type II ghost factors occur when the terms oscillate within a smaller subset on the right half of the complex plane. An example is with $N = 3\times7\times83\times151$ and $l = 12$. The reduced fraction of $N/l$ gives $p = 3$ and $q = 4$, which alternates between $1$ and $i$ in the complex plane, so the measurement outcome oscillates between $1$ and $0.5$ for each $k$, resulting in ${\Pr}(m) \approx 0.75$ overall.

Another example, with the same $N$ and $l = 15$, $35$, $105$, $1245$, gives the reduced fraction with coprimes $q=5$ and $p=1$, $4$, $3$, $2$. The $p=1$ branch always ends up being on the right side of the plane as the first step involves a single rotation $2\pi/q$ away from the starting point. Hence, the $p=1$ branch gives the worst ghost factor for a given $q$.

The sum of the form $\sum_{m=0}^{q-1} e^{-i2\pi m^2/q}$ has been previously worked out by Gauss \cite{Berry1988,Gauss1808}, and by taking the real component of the result, we find
\begin{align}\label{eq:type-II-sum}
  \ev{\Pr(p=1,q)}^{(q-1)} &= \frac{1}{2}\pqty{
    1 + \frac{1}{q}\sum_{m=0}^{q-1}\cos(\frac{2\pi m^2}{q})
  } \\
  &=
  \begin{cases}
    \frac{1}{2}\pqty{1+\frac{1}{\sqrt{q}}} & \text{if}~q\bmod4 = 0,1;\\
    0 & \text{otherwise}.
  \end{cases}\nonumber
\end{align}
A trial factor meets this condition when it is of the form $l = (4k+\lambda) \times n$ where $\lambda = 0,1$, integer $k\geq 1$, and $n$ divides $N$.

From Eq.~\eqref{eq:type-II-sum}, Type II ghost factors get smaller with larger $q$. Note that this involves the full sum from $m=0$ to $q-1$. However, since the worst-case scenarios occur at small $q$, the problematic factors of interest will have gone through several complete cycles after the lower limit of $M$ have already been met, so the final average $\ev{\Pr(p=1,q)}$ will be approximated by this sum.

We reiterate that Type II ghost factors cannot be improved with large $M$, since increasing the number of pulses only duplicates the above sum several times with some extra terms coming from incomplete cycles of $(0,q-1)$. In fact, as $M\to\infty$, the contributions of these extra terms become negligible, and $\ev{\Pr(p=1,q)}$ becomes \emph{exactly} the value given by Eq.~\eqref{eq:type-II-sum}.

A consequence of the above is that the Type II ghost factors limit the ability for us to discern between factors and nonfactors. The presence of $q=4$ ghost factors mean that $\ev{\Pr({\scriptstyle\text{factor}})}-\ev{\Pr({\scriptstyle\text{nonfactor}})} \leq \frac{1}{2}(1-{1}/{\sqrt{4}}) = 1/4$, which can be a narrow margin in the presence of noise. This value also determines where we would place the cutoff between factors and nonfactors, as setting it too low (and close to the worst nonfactor) would cause us to misidentify nonfactors as factors, while the converse would cause us to misidentify factors as nonfactors.

Furthermore, this difference becomes zero after a long time due to decoherence, whether it is because both states relax to the ground state or because both dephase to a completely mixed state. As such, the study of this worst-case scenario is an important practical consideration when implementing Gauss sum factorization experimentally. To that end, we will take a closer look at this in the next section.

\section{Results}
\subsection{Discernability\label{sec:discernability}}
As we have seen, the signal difference between factors and nonfactors is bounded above by Type II ghost factors. Practically speaking, given the result of a Gauss sum factorization experiment, we wish to choose a $\ev{\Pr}_{\text{cutoff}}$ to identify all $l$ such that $\ev{\Pr(l)} > \ev{\Pr}_{\text{cutoff}}$ to be factors, and all $l$ such that $\ev{\Pr(l)} < \ev{\Pr}_{\text{cutoff}}$ to be nonfactors. Previous work suggests the use of the worst nonfactor as the cutoff \cite{Ng2010}, which is $\ev{\Pr}_{\text{cutoff}} = 0.75$ in our case, but this does not take into account decoherence, which might cause the signal of a factor to fall below that threshold.

\begin{figure}
  \centering
  \hspace*{-2mm}\includegraphics{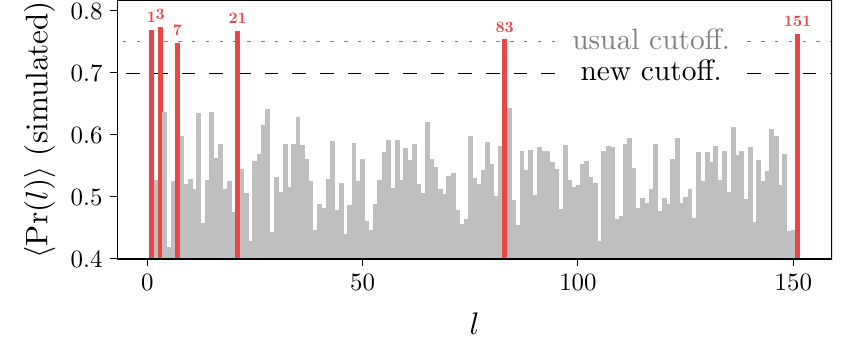}
  \caption{\label{fig:cutoff-example}Simulated example of Gauss sum with $T_2 = 500~\mathrm{ns}$ and Gaussian random measurement noise added to Eq.~\eqref{eq:noisy-gauss-sum-br}. Factors are labelled above the corresponding bars. If the usual cutoff of $3/4$ is chosen, some of the factors, due to decoherence causing them to fall below the chosen cutoff, might be misidentified as nonfactors. We suggest a new cutoff based on the expected decay of the trial factors, and choosing it to be between the expected values of the factor and the worst ghost factor. This provides leeway for fluctuations without misidentification of the factors as nonfactors or vice versa.}
\end{figure}

Instead, an improved cutoff would be halfway between the expected signal of the factor and the worst nonfactor
\begin{equation}\label{eq:cutoff}
\ev{\Pr}_{\text{cutoff}} = \frac{1}{2}\pqty{\widetilde{c}^{(m)}_N({\scriptstyle\text{worst nonfactor}}) + \widetilde{c}^{(m)}_N({\scriptstyle\text{factor}})}.
\end{equation}
This not only accounts for the decoherence of the qubit, but also allows for some leeway for fluctuations and measurement errors. An illustration of this is shown in  Fig.~\ref{fig:cutoff-example}. We multiplied the value of each Gauss sum term with a random number sampled from a normal distribution with a standard deviation of 0.04 to simulate the presence of measurement noise. The cutoff calculated using Eqs.~\eqref{eq:noisy-gauss-sum-br} \& \eqref{eq:cutoff} is placed where the signal of the factors can fluctuate without crossing over the threshold.

To study this in more detail, we define the \emph{discernability} of an experimental implementation of a Gauss sum factorization
\begin{equation}\label{eq:discernability-definition}
  \begin{aligned}
    \mathcal{D}(M,N)
    &\equiv 2\bigg(\ev{\Pr({\scriptstyle\text{factor}},N)}^{(M)} \\
    &\qquad-\ev{\Pr({\scriptstyle\text{worst ghost factor}},N)}^{(M)}\bigg)\\
    &= \widetilde{C}^{M}_N({\scriptstyle\text{factor}})-\widetilde{C}^{M}_N({\scriptstyle\text{worst ghost factor}}).
  \end{aligned}
\end{equation}
At this point, the definition does not prescribe the worst ghost factor, to allow for experimenters to skip past known nonfactors to improve the signal. This can be done by preprocessing the number to be factorized, which will be covered in Sec.~\ref{sec:preprocessing}.

The factor of two sets the discernability within the range $[0,1]$ when factors can be discerned from nonfactors. The former holds when the qubit completely decoheres, while the latter holds when all ghost factors are eliminated, leaving only ``well-behaved'' ghost factors such that $\mathcal{D} = 2\pqty{1-0.5} = 1$. Meanwhile, $\mathcal{D} < 0$ means that the signal of the nonfactor exceeds the factor, so a misidentification will occur with certainty.

We juxtapose this to the \emph{contrast} $\mathcal{V}$ used in other Gauss sum factorization experiments \cite{Gilowski2008,Mehring2007}, which is an adaptation of the Michelson contrast \cite{Born1999} to the Gauss sum factorization scheme. Contrast is given by
\begin{equation}
\begin{aligned}\label{eq:contrast}
    \mathcal{V} &= \frac{I_{\text{max}}-I_{\text{min}}}{I_{\text{max}}+I_{\text{min}}}
    = \frac{1-a}{1+a},\\
    a &= \frac{1}{{\scriptstyle\text{no. of nonfactors}}}\sum_{\text{nonfactors } l} \abs{C^{(M)}_N(l)}\\
    &= \frac{1}{{\scriptstyle\text{no. of nonfactors}}}\sum_{\text{nonfactors } l} \abs{2\ev{\Pr(l,N)}^{(M)} - 1}
\end{aligned}
\end{equation}
with $a$, the average over the absolute values of the Gauss sums of nonfactors, playing the role of the minimum intensity. The contrast reflects the overall performance of the system, and is a useful gauge of the effectiveness of a particular Gauss sum factorization.

On the other hand, the discernability quantifies the system at its worst behavior. It does not supplant contrast as a descriptor of the system's overall performance, but rather serves to indicate the limits of the Gauss sum factorization where noise might render the scheme ineffectual. This is useful as an indicator when pushing a system to its decoherence limits, while achieving a target discernability to maintain the effectiveness of the factorization scheme.

We apply the results from the Bloch-Redfield equation to investigate the behavior of discernability in the presence of noise. If all trial factors are present, the worst Type II ghost factor occurs for $q=4$, and the summands of the factor and ghost factor follow a $T_2$ decay given by
\begin{align}
\widetilde{c}^{(m)}_N({\scriptstyle\text{factor}}) &= \widetilde{c}^{(m)}_N(q=1)\\
&= e^{-(m+1)\Gamma_2(\tau+t_\pi)}\nonumber\\
\widetilde{c}^{(m)}_N({\scriptstyle\text{worst ghost factor}}) 
&= \widetilde{c}^{(m)}_N(q=4) \\
&= \begin{cases}
e^{-(m+1)\Gamma_2(\tau+t_\pi)} & \text{for even $m$};\nonumber\\
0 & \text{for odd $m$}.
\end{cases}
\end{align}
This gives the Gauss sums
\begin{align}
\widetilde{C}^{(M)}_N(q=1) &=\frac{1}{M+1} \sum_{m=0}^M e^{-(m+1)\Gamma_2(\tau+t_\pi)}\nonumber\\
&= \frac{1-e^{-(M+1)\Gamma_2(\tau+t_\pi)}}{(M+1)(e^{\Gamma_2(\tau+t_\pi)}-1)},\label{eq:noisy-gauss-factor}\\
\widetilde{C}^{(M)}_N(q=4) &=\frac{1}{M+1} \sum_{m=0}^{\lfloor M/2 \rfloor} e^{-(2m+1)\Gamma_2(\tau+t_\pi)}\nonumber\\
&= \frac{1-e^{-(2\lfloor M/2 \rfloor+2)\Gamma_2(\tau+t_\pi)}}{(M+1)(e^{\Gamma_2(\tau+t_\pi)}-e^{-\Gamma_2(\tau+t_\pi)})}\label{eq:noisy-gauss-nonfactor}\\
&\approx \frac{1-e^{-(M+2)\Gamma_2(\tau+t_\pi)}}{(M+1)(e^{\Gamma_2(\tau+t_\pi)}-e^{-\Gamma_2(\tau+t_\pi)})}\nonumber.
\end{align}
In the last step, we approximate $\lfloor M/2\rfloor \approx M/2$ as $M$ is generally large when the lower limit for the $M$ for suppressing Type I ghost factor is met. The discernability is the difference, given by
\begin{equation}
  \mathcal{D}(M,N) = \frac{1-e^{-M\Gamma_2\pqty{\tau+t_\pi}}}{\pqty{M+1}\pqty{e^{2\Gamma_2\pqty{\tau+t_\pi}}-1}}\label{eq:discernability}.
\end{equation}
At large $M$, $\mathcal{D}\to0$, which is the statement that it gets increasingly difficult to discern the factors and nonfactors as the number of pulses increase.

\begin{figure}
  \centering
  \hspace*{-2mm}\includegraphics{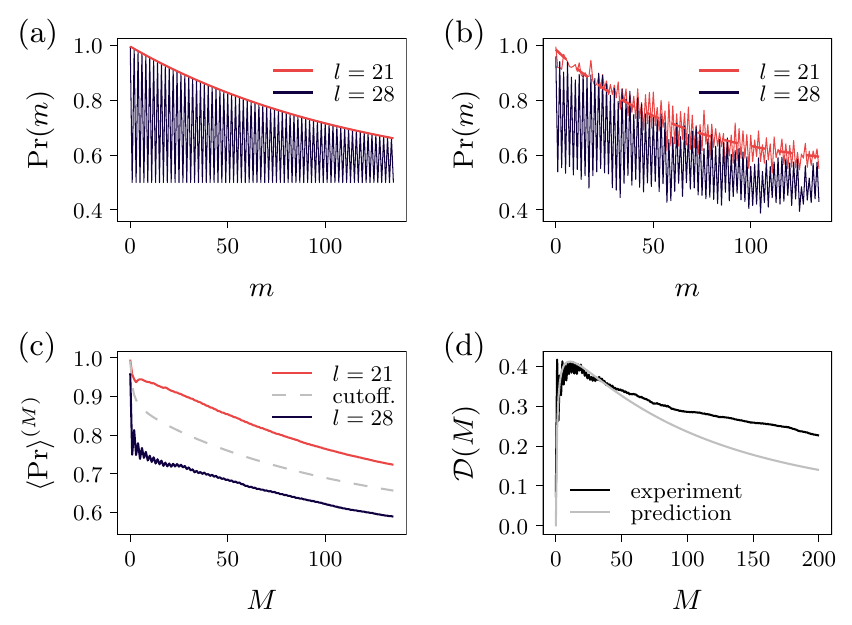}
  \caption{\label{fig:decay-cutoff-dis}Gauss summand decay behavior, cutoffs, and discernability. (a) Expected behavior of Gauss summands from theory. The signal from the factor ($l=21$, upper solid line) decays exponentially with a characteristic time $T_2$, while the signal from the ghost factor ($l=28$, lower solid line) oscillates between within the envelope formed by the factor. (b) Experimental measurements of signals from the factor (upper solid line) and ghost factor (lower solid line, oscillating). The dashed line is an exponential fit to yield $T_2$. (c) Decay behavior of the Gauss sum of the factor (upper solid line) and nonfactor (lower solid line), together with the cutoff (dashed line) calculated from the measured $T_2$. Our suggested cutoff always lie between the factor and the ghost factor, so misidentification is unlikely as long as the fluctuations are smaller than the separation between the two. (d) Experimental (black) and theoretical (gray) discernability of the two factors. At very small $M$ the nonfactor is not sufficiently suppressed while at large $M$ decoherence closes the gap between the factor and the nonfactor, so both extremes result in a reduced discernability.}
\end{figure}

The discernability of a system can be thought of as the tolerance of the Gauss sum factorization to measurement uncertainty. This is illustrated in Fig.~\ref{fig:decay-cutoff-dis}, where we have plotted the experimental behavior of a factor and a ghost factor with the cutoff calculated with Eq.~\eqref{eq:cutoff}, and the experimental and theoretical discernability from Eq.~\eqref{eq:discernability}. We applied a long sequence of pulses for a factor ($l=21$) and the worst ghost factor ($l=28$, which has $q=4$) with pulse delay $\tau=30~\mathrm{ns}$ and $\pi$-pulse time $t_\pi=25~\mathrm{ns}$. We performed a fit shown in Fig.~\ref{fig:decay-cutoff-dis}(b) to find $T_2 = 3.6 \pm 0.5~\mu\text{s}$. Graphically, the $\mathcal{D}$ plotted in Fig.~\ref{fig:decay-cutoff-dis}(d) is twice the gap between the factors (upper solid line) and nonfactors (lower solid line) shown in Fig.~\ref{fig:decay-cutoff-dis}(c). The discernability is reduced at very small $M$, where the nonfactor is not yet sufficiently suppressed, and at large $M$, where the decoherence of both the factor and nonfactor causes the separation between the two to close. We can tolerate a much smaller discernability in a system where measurement uncertainty is low, but a larger discernability would be needed in a noisy system for a bigger allowance for fluctuations without the misidentification of factors or nonfactors.

In that sense, for a target $\mathcal{D}_\text{target}$, we require $M < M_{\text{max}}$, with $M_{\text{max}}$ given by inverting Eq.~\eqref{eq:discernability} and taking the principal branch of the Lambert W function to find
\begin{equation}\label{eq:upper-limit}
  M_{\text{max}} = M_0 W\pqty{-\frac{e^{-\frac{1}{M_0}\pqty{\frac{1}{\mu\mathcal{D}_\text{target}}-1}}}{M_0\mu\mathcal{D}_\text{target}}} + \frac{1}{\mu\mathcal{D}_\text{target}} - 1,
\end{equation}
where $\mu=e^{\frac{2}{M_0}}-1$ and $M_0 = T_2/(\tau+t_\pi)$. $M_0$, the ratio between the coherence time and pulse duration, is a naive estimate of the maximum circuit depth.

Hence, Type II ghost factors provide an upper bound for $M$ by taking into account the decreasing discernability with an increasing number of pulses. Together with a lower bound due to Type I ghost factors, this sets a limit to the largest factorizable number at $\sqrt[4]{N/4} \sim M_{\text{max}}$.

There are two options to increase the largest factorizable number: increase the discernability of the system so that we can achieve $\mathcal{D}_{\text{target}}$ with a larger $M$, or increase the ratio $T_2/(\tau + t_\pi)$. For the former, we introduce the technique of preprocessing in Sec.~\ref{sec:preprocessing}. For the latter, this involves fitting a larger number of pulses within the coherence time of the qubit, which we explore in Sec.~\ref{sec:fit-more-pulse}.

\subsection{\label{sec:preprocessing}Preprocessing}
From Eq.~\eqref{eq:type-II-sum}, the two worst ghost factors occur when the trial factors are certain multiples of 4 or 5. Hence, if we avoid trial factors that are multiples of either 4 or 5, we can eliminate these ghost factors and increase the discernability. To do so, we can reduce $N$ to another number $N^{(2)}$ does not have 4 or 5 as a factor. This is easily done as the divisibility tests for 2 and 5 are simple, and the integer division of a number is computationally cheap.

The exact steps are as follows:
\begin{enumerate}
    \item If $N$ is even, divide $N$ by 2 until an odd number $N^{(1)}$ is obtained. Store the number of divisions as $n_2$.
    \item If the last digit of $N^{(1)}$ is 5, divide $N^{(1)}$ by 5 until a number $N^{(2)}$ without 5 as the last digit is obtained. Store the number of divisions as $n_5$.
    \item Perform Gauss sum factorization on $N^{(2)}$ with odd trial factors that are not multiples of 5. The prime decomposition of $N$ is $2^{n_2} \times 5^{n_5}\times\{\text{prime decomposition of $N^{(2)}$}\}$.
\end{enumerate}
With this approach, the worst remaining Type II ghost factor is at $q=9$, which brings the upper limit of the discernability up to $\mathcal{D} \leq (1-1/\sqrt{9}) = 0.67$, an improvement over the unprocessed case $\mathcal{D} \leq 0.5$. This improvement can make it far easier to discern between factors and nonfactors, especially when the total time of the pulse train approaches the decoherence limit of the qubit.

\begin{figure}
  \centering    
  \hspace*{-2mm}\includegraphics{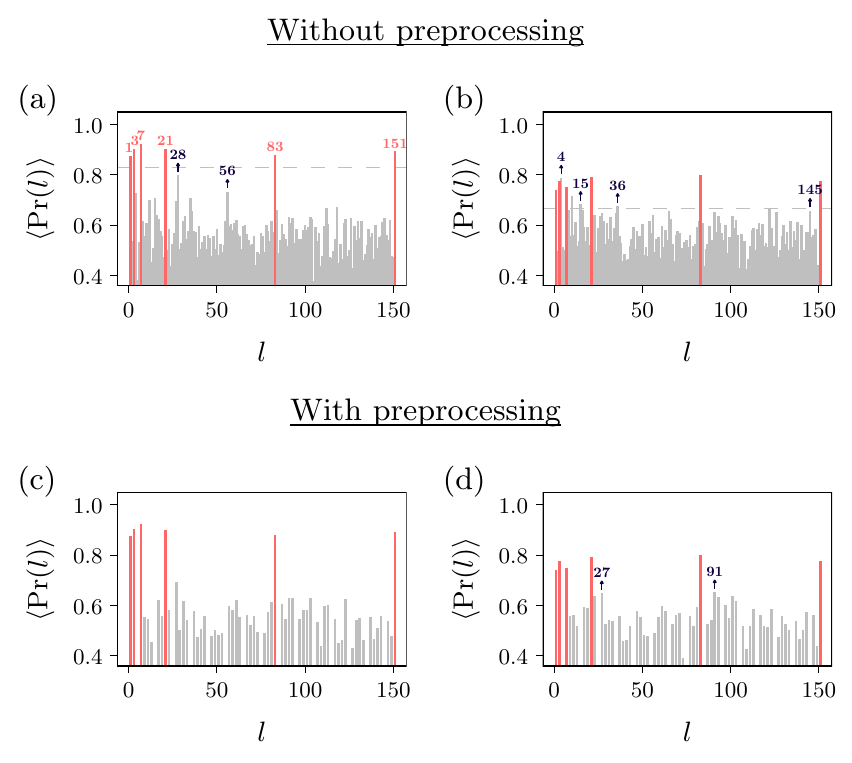}
  \caption{\label{fig:experiment-processing}Experimental results for the Gauss sum factorization of $263193 = 3\times 7 \times 83 \times 151$ with the total number of pulses $M=17$. The factors are labelled above the corresponding bar without arrows, and the cutoff from Eq.~\eqref{eq:cutoff} is marked out with a dashed line. The qubit drive is on resonance in (a \& c) and with detuning noise in (b \& d). The noise is introduced by randomly detuning the qubit drive to $\omega_d = \omega_q - \delta$, where $\delta$ is sampled from a uniform distribution in the range $[0,2\pi\times250~\mathrm{MHz}]$. The results are shown without preprocessing (a \& b) and with preprocessing (c \& d). In (a), ghost factors $l=28$ and $l=56$, labelled with arrows, are close to the cutoff, although there is no misidentification of factors and nonfactors. In (b), the ghost factor $l=4$ exceeds some of the factors, so there will be misidentification regardless of how the cutoff is adjusted. In addition, the ghost factors $l=15$ and $l=36$, labelled with arrows, exceed the chosen cutoff, so there is a risk of misidentifying these nonfactors depending on the choice of cutoff. In (c \& d), there is a clear separation between factors and nonfactors, and there is no longer any misidentification of either. The worst nonfactors after preprocessing are $l=27$ and $l=91$, labelled with arrows, which are the next worst ghost factors with $q=9$ and $q=13$}
\end{figure}

We consider such a situation in our experimental setup. First, we performed the Gauss sum factorization with $M=17$, with and without preprocessing, and the qubit and driving frequency on resonance. On resonance, the coherence time is $T_2 = 3.5 \pm 0.5~\mu\text{s}$, and the results are shown in Fig.~\ref{fig:experiment-processing}a. Then, we intentionally introduced noise by randomly detuning the driving frequency of the qubit away from resonance, such that $\omega_d = \omega_q - \delta$, where $\delta$ is sampled from a uniform distribution with $\delta \in [0,2\pi\times250~\mathrm{MHz}]$, reducing the coherence time of the system to $T_2 = 0.4 \pm 0.2~\mu\text{s}$. Hence, when noise is intentionally introduced, the qubit is operating in a region where the total experiment time $17\times(\tau+t_\pi) = 0.94~\mu\text{s}$ \emph{exceeds} the coherence time. The results of the qubit operated at this region is shown in Fig.~\ref{fig:experiment-processing}b.

Type II ghost factors like those at $l=4$ and $l=36$ ($q=4$),  $l=15$ and $l=145$ ($q=5$) are eliminated, leaving a clear separation between the actual factors and the nonfactors. Quantitatively, the discernability increases from $\mathcal{D} = 0.198 \pm 0.006$ to $\mathcal{D} = 0.402 \pm 0.006$ on resonance, and $\mathcal{D} = -0.032 \pm 0.008$ to $\mathcal{D} = 0.240 \pm 0.008$ detuned, where the uncertainties are calculated by propagating the standard error of the signal of the factors through Eq.~\eqref{eq:discernability-definition}. The drastic improvement in discernability can be verified visually when comparing panels (a) with (c), and (b) with (d), in Fig.~\ref{fig:experiment-processing}. Note also that $\mathcal{D} < 0$ before preprocessing, which is because the nonfactor 4 will be misidentified as a factor no matter the choice of cutoff. This shows that preprocessing can improve the utility of the Gauss sum factorization scheme, allowing it to be used even when the unmodified version would result in an incorrect result.

In addition, preprocessing also improves the overall contrast of the system. The on-resonance contrast increases from $\mathcal{V} = 0.750\pm0.010$ to $\mathcal{V} = 0.780 \pm 0.012$, while the detuned contrast increases from $\mathcal{V} = 0.765 \pm 0.009$ to $\mathcal{V} = 0.792 \pm 0.012$, where the uncertainties are calculated by propagating the standard error of the signal of the nonfactors through Eq.~\eqref{eq:contrast}. We note that the detuned contrast is larger than the on-resonance contrast, which makes it seem like detuned case provides a better separation between the factors and the nonfactors. This is not the case, but is rather due to the definition of the contrast in Eq.~\eqref{eq:contrast}, where the ``maximum intensity'' is taken to be 1, which does not account for decoherence of the factors. This is easily fixed by replacing 1 with the experimentally-measured signals of the factors, but we elected to keep the definition consistent with the literature \cite{Gilowski2008,Mehring2007}.

While the increase of both figures of merit reflect the improved separation between the factors and the nonfactors, we would not have guessed the presence of the misidentified factor from the reported contrast. This is however reflected by the negative discernability of the detuned case before preprocessing, which reiterates the differing purposes of either figures of merit. Here, the increased $\mathcal{D}$ means that we can increase the maximum $M$, so preprocessing can increase the upper limit of $N$ that can be factorized with Gauss sums.

\subsection{\label{sec:fit-more-pulse}Fitting More Pulses Within Coherence Time}
Another method to increase the largest possible $N$ is to increase the ratio $T_2/(t_\pi+\tau)$, which corresponds to fitting more pulses within the $T_2$ time of the qubit. One approach is to increase $T_2$ or decrease $t_\pi$, which is mostly an engineering challenge with regards to the the experimental equipment used. On the other hand, we can also decrease the delay $\tau$ between successive pulses. However, in doing so, we observed a behavior that will require an extension of the noise model beyond the Bloch-Redfield theory.

\subsubsection{\label{sec:decrease-pulse-delay}Decreasing Pulse Delay}
In the previous sections, we have performed the pulse sequences with a delay $\tau = 30~\mathrm{ns}$, which is in the order of the $\pi$-pulse time $t_\pi = 25~\mathrm{ns}$. The results in Fig.~\ref{fig:decay-cutoff-dis} show that the experiments agree with the theoretical predictions, and the system does dephase with a characteristic time $T_2$ independent of the pulse sequence.

\begin{figure}
  \centering
  \hspace*{-2mm}\includegraphics{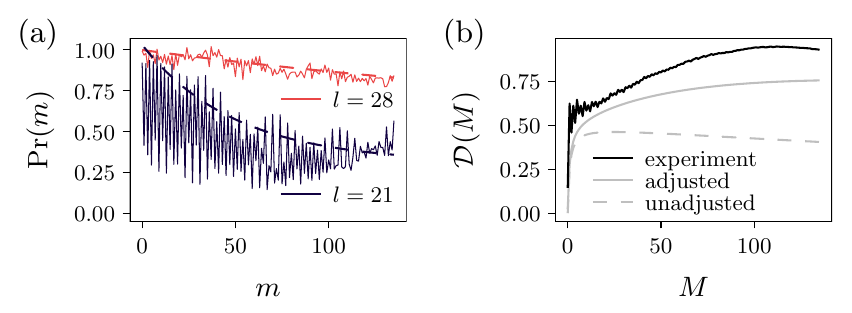}
  \caption{\label{fig:decay-cutoff-dis-small-delay}Experimental Gauss summand and discernability for factor ($l=21$, upper lines) and nonfactor ($l=28$, lower lines) (similar to Figs.~\ref{fig:decay-cutoff-dis}(b \& d)) with pulse delay $\tau = 1~\mathrm{ns} \ll t_\pi$. The dashed lines are the fits used to find the $T_2$ times. We found that in this regime, the $T_2$ time of the factor was much larger than the $T_2$ time of the nonfactor. This required an adjustment of the discernability, which is shown in (b). The adjusted discernability agrees better with the experimental results.}
\end{figure}

However, when performing the pulse sequence with a much shorter pulse delay, we found the qubit to decay at a different rates: $T_2^{(1)}$ for the factors and $T_2^{(4)}$ for the $q=4$ ghost factor with $T_2^{(4)} < T_2^{(1)}$. This is shown in Fig.~\ref{fig:decay-cutoff-dis-small-delay}(a). We repeated the procedure performed in Fig.~\ref{fig:decay-cutoff-dis}(b) but with $\tau=1~\mathrm{ns}$, and fitted the graph to find $T_2^{(1)} = 8.2\pm0.5~\mu\text{s}$ for the factor and $T_2^{(4)} = 1.2\pm0.3~\mu\text{s}$ for the nonfactor.

In lieu of these findings, the discernability was adjusted for the two $T_2$ times as
\begin{equation}\label{eq:adjusted-discernability}
    \mathcal{D}(M,N) = 
    \left.\widetilde{C}^{M}_N(q=1)\right\rvert_{T_2^{(1)}}
    - \left.\widetilde{C}^{M}_N(q=4)\right\rvert_{T_2^{(4)}},
\end{equation}
where $\widetilde{C}^{M}_N$ for both cases are given by Eqs.~\eqref{eq:noisy-gauss-factor} \& \eqref{eq:noisy-gauss-nonfactor}.

When the decay times are different, there are no equivalent closed-form expressions for Eqs.~\eqref{eq:discernability} \& \eqref{eq:upper-limit}, but they can be calculated numerically using the same procedure. This is done in Fig.~\ref{fig:decay-cutoff-dis-small-delay}(b), where we compare the adjusted discernability in Eq.~\eqref{eq:adjusted-discernability} against the unadjusted discernability in Eq.~\eqref{eq:discernability} with the averaged decay rate $T_2^{(\text{avg.})} = (T_2^{(1)} +  T_2^{(4)})/2$. We found Eq.~\eqref{eq:adjusted-discernability} to better reflect the behavior of the experimental data.

\subsubsection{Filter Function Formalism\label{sec:filter-function}}
Where we have made a phenomenological adjustment to the discernability from our observation that $T_2$ is different for factors over ghost factors, we also set out to understand why there is such a difference in decoherence behaviors. We offer some preliminary qualitative explanations by using the filter function approach, more commonly seen the study of quantum control in the presence of universal noise \cite{Green2013}.

In the filter function formalism, qubit dephasing is described by the factor $e^{-\chi(t)}$, where $\chi(t)$ is the coherence integral given by \cite{Green2013}
\begin{equation}\label{eq:decay-calculation-filter-function}
\chi(t) = \frac{1}{\pi} \sum_{ij}\int_{-\infty}^{\infty}\dd{\omega} S_{ij}(\omega)g_{ij}(\omega;t),
\end{equation}
where $g_{ij}(\omega;t)$ is the filter function, so named as it filters out the power spectral density $S_{ij}(\omega)$ of the noise present in the $i,j = \{x,y,z\}$ component of the qubit control. $g_{ij}$ is purely determined by the control pulses
\begin{align}\label{eq:filter-function}
g_{ij}(\omega) &= \frac{1}{\omega^2}\bqty{ R_\omega(\omega)R_\omega^\dagger(\omega)}_{ij},\\\label{eq:rotation-frequency-domain}
R_\omega(\omega) &= -i\omega\int_0^t\dd{t'} R^{(\text{ctrl})}(t') \, e^{i\omega t'}.
\end{align}
Here, $R^{(\text{ctrl})}$ is the control unitary written as a 3-by-3 rotation matrix.

\begin{figure}
  \centering
  \hspace{-.25em}\includegraphics{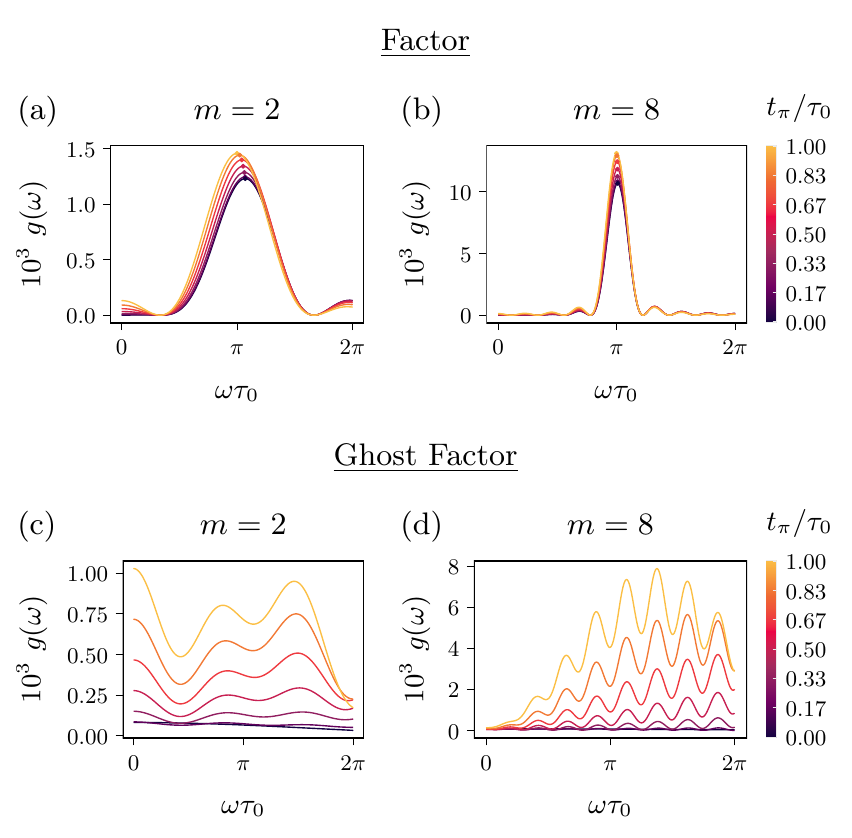}
  \caption{Filter functions of a factor (a \& b) and the worst ghost factor (c \& d), with constant $\tau_0 = \tau + t_\pi$. For all four subfigures, the values of $t_\pi/\tau_0$ of each line is marked out on the color scale in the same order, with $t_\pi/\tau_0=1$ for the uppermost lines and $t_\pi/\tau_0=0$ for the lowermost lines. For the factor, assumption of instantaneous pulses (here $t_\pi=0$) in noise spectroscopy is lifted, resulting in some deviations to the peak height and location. For the ghost factor, the pulse sequence filters out a large range of frequencies of the noise spectrum.}\label{fig:filter-functions}
\end{figure}

In this context, the control pulses specify a filter function in the frequency domain, which filters out certain frequency ranges of the noise. Since the control pulses are determined by the trial factors in the Gauss sum calculations, there is a corresponding filter function for each trial factor. They are worked out in Appendix \ref{apd:gauss-sum-filter-function}, and the filter functions for a factor and the worst ghost factor are shown in Fig.~\ref{fig:filter-functions}.

Two comments are in order. Firstly, our pulses sequences are built as a spin-echo-like pulse train, initially with the intention to remove the decoherence effect of detuning within the Bloch-Redfield model (see Eq.~\eqref{eq:bloch-redfield-decoherence}). As such, the pulse sequence for a factor happens to be a Carr-Purcell-Meiboom-Gill (CPMG) pulse train, which in the limit of instantaneous pulses results in a Dirac-delta approximation $\propto\operatorname{sinc}(\omega-\frac{\pi}{\tau})\approx \delta(\omega-\frac{\pi}{\tau})$ in the frequency domain centred around a target frequency that depends on the delay time between each pulse \cite{Bylander2011}. The instantaneous pulse assumption approximately holds when $\tau\gg t_\pi$.

The benefit of the filter function approach is that we are able to incorporate finite pulse time effects so that the delay time between each pulse can be reduced. This allows us to fit more operations within the coherence time of the qubit, like in our experimental implementation in the preceding discussion, where $\tau \ll t_\pi$. As we can work out the filter function without the instantaneous pulse assumption, we can predict the region of the noise spectrum that will contribute to the decoherence of the qubit. Importantly, we see that the peak of the filter function is roughly the same as in the instantaneous case, so long as we calculate the filter function using $\tau + t_\pi$ for the ``pulse delay'', that is, $ \delta(\omega-\frac{\pi}{\tau}) \to \delta(\omega-\frac{\pi}{\tau+t_\pi})$. Increasing the number of pulses $m$ still brings the filter function closer to the Dirac delta function, as it does in noise spectroscopy.

Secondly, the filter function caused by the $q=4$ ghost factor spreads out over a wide range of frequencies. As the filter function is integrated over the spectrum, a broad portion of the environmental noise will contribute to the decoherence of the qubit. For this reason, it is likely that the qubit will decohere faster for a ghost factor, compared to a factor, where only a narrow region contributes. Whether or not this is true depends upon the actual noise spectrum.

\begin{figure}
  \centering
  \hspace{-1em}\includegraphics{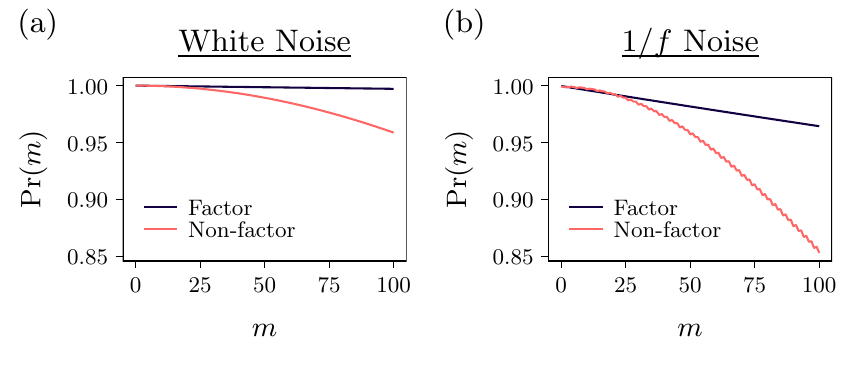}\vspace{-2ex}
  \caption{Decoherence of factor and worst ghost factor in the filter function formalism in the presence of white noise (a) and $1/f$ noise (b). The dephasing behavior is found by using Eqs.~\eqref{eq:decay-calculation-filter-function}~to~\eqref{eq:rotation-frequency-domain} with $S(\omega)=\text{constant}$ for white noise and  $S(\omega) \propto 1/\omega$ for $1/f$ noise. In both cases, the nonfactor (lower lines) decays more rapidly than the factor (upper lines). Note that only the envelope is plotted, so the rapid oscillation of the nonfactor is not shown.}\label{fig:decay-from-filter-function}
\end{figure}

For a sense of how this might be so, we use $S(\omega) = \text{constant}$ (white noise) and $S(\omega) \propto 1/\omega$ (1/f noise) to calculate Eq.~\eqref{eq:decay-calculation-filter-function}. These noise statistics are ubiquitous in electronics, and their presence and mechanism in superconducting qubits have been widely studied \cite{Paladino2002, Koch2007}. The constants and proportionality factors are chosen so that the standard deviation for both noises are the same, but otherwise arbitrary, as we only wish to explore the qualitative properties of the filter functions shown in Fig.~\ref{fig:filter-functions}.

The resulting behavior is plotted in Fig.~\ref{fig:decay-from-filter-function}. For both types of noise, the nonfactor always decays faster. This could be an explanation for the phenomenological adjustment made in Eq.~\eqref{eq:adjusted-discernability}, where the $T_2$ time of a factor is an order of magnitude larger than worst ghost factor.

Undoubtedly, a quantitative comparison of this decoherence model with the experimental data can only be done with knowledge of the spectrum of the environmental noise, which is beyond the scope of this experiment. A possible approach to this is offered in Sec.~\ref{sec:noise-spectroscopy}.

\section{Discussion}
\subsection{Type II ghost factors}
We have shown that Type II ghost factors can affect a Gauss sum factorization computation even in regimes where Type I ghost factors might be suppressed. This is a drawback to the scheme, especially when pushing the limits of the number of digits of $N$ to be factorized.

The effect of the Type II ghost factor can be characterized by the discernability $\mathcal{D}$, which can be found after measuring the $T_2$ time of the system when performing the pulse sequence for just the factor if the pulse delay is not much shorter than the $\pi$-pulse time. This determines the maximum number of pulses $M$ possible by requiring that the measurement errors should be smaller than $\mathcal{D}$, which in turn sets a limit on the maximum factorizable number $N$.

As discernability is a figure of merit tailored to Gauss sum factorization, a better estimate of this upper limit can be obtained, given that the tolerances of the experimental setup are known. For example, in the setup used in Fig.~\ref{fig:decay-cutoff-dis}, the maximum number of pulses without considering the specifics of the Gauss sum factorization scheme is given by $M_0 = T_2/(\tau+t_\pi) = 56$. The lower bound of the coherence time is taken to provide a conservative estimate. Together with the lower limit $\sqrt[4]{N/4} < M_0$, we find $\log_{10}(N) < 7.6$. Meanwhile, from the residual standard error in Fig.~\ref{fig:decay-cutoff-dis}b, the uncertainty in the signal is $\delta\Pr = 0.06$. That means that $\mathcal{D}_{\text{target}} = 0.12$ would be sufficient to separate factors from nonfactors with certainty. From Eq.~\eqref{eq:upper-limit}, we find $M_{\text{max}} = 225$. With $\sqrt[4]{N/4} < M_{\text{max}}$, this implies $\log_{10}(N) < 10.0$.

Therefore, by incorporating the tolerances of the setup, the experimentalist can perform more pulses to factorize 10-digit numbers instead of 8-digit numbers, which is a large improvement. Importantly, because discernability is operationally twice the signal difference between the worst nonfactor and the factors, the chosen target discernability guarantees that they are operating in a regime where the factorization scheme is still effective. Furthermore, with preprocessing, which increases the discernability for a given coherence time, the maximum number of pulses can be further increased.

The same procedure can also be done when the pulse delay is much shorter than the $\pi$-pulse time, except with a phenomenological adjustment that requires the measurement of $T_2$ of the worst ghost factor. While it seems problematic that we can only predict the decay behavior of a trial factor after actually performing the measurement to obtain its coherence time, this worry can be alleviated with Eq.~\eqref{eq:gauss-sum-coprime}, which shows that the \emph{property} of the trial factor in relation to $N$ --- which category it belongs to and the resulting coprimes $p$ and $q$ --- determines the resulting filter function of the pulse sequences. Hence, we can measure the $T_2$ time of all factors by measuring the $T_2$ time for a known factor ($l=1$), and likewise for the worst Type II ghost factor by measuring the system using a smaller $N$ and a known Type II ghost factor. These $T_2$ times can be fed back into Eqs.~\eqref{eq:cutoff} \& \eqref{eq:adjusted-discernability} to calculate the discernability and the cutoff. The discernability can be inverted to work out the upper limit for the number of pulses, and hence, the largest possible $N$.

\subsection{Preprocessing}
A strategy to improve the discernability would be to reduce $N$ to another number $N^{(k)}$ that is coprime to the worst few Type II ghost factors. In Sec.~\ref{sec:preprocessing}, we provided the steps to remove the first two ghost factors to improve the discernability from to $\mathcal{D} \leq 0.5$ to $\mathcal{D} \leq 0.67$.

This method is easily extendable. For example, the next worst ghost factor $q=9$ can be removed by performing a division test for 9 (by checking if the sum of digits of $N^{(2)}$ is a multiple of 9), and dividing $N^{(2)}$ by 9 until the division test fails to obtain $N^{(3)}$. The number of divisions by 9 is stored as $n_9$. Finally, Gauss sum factorization is performed on $N^{(3)}$ with trial odd trial factors that are neither multiples of 5 nor 9. The prime decomposition of $N$ is $N = 3^{2n_9+1}\times\dots$ if $N^{(3)}$ is a factor of $3$ and $N = 3^{2n_9}\times\dots$ otherwise. This would improve the discernability further to $\mathcal{D} \leq 0.72$.

\subsection{Noise Spectroscopy\label{sec:noise-spectroscopy}}
The filter function formalism potentially explains the differing $T_2$ times between factors and nonfactors in the presence of common types of environmental noise when shortening the pulse delay to a value much smaller than the pulse time. However, this claim has to be checked with experiment, which can only be done if we can reconstruct the actual noise spectrum of the environment.

One possibility to do so is to perform spectroscopy experiments on the qubit to measure the spectrum of the environmental noise, which will allow us to predict the decay of the nonfactors by numerically integrating Eq.~\eqref{eq:decay-calculation-filter-function} to find the associated $T_2$ times. This will lead us to a better estimate of the discernability, and predict the behavior of the qubit at large $m$, which is required when factoring larger numbers.

\section{Conclusion}
We investigated the effects of decoherence and ghost factors in a Gauss sum factorization scheme. We found that Type II ghost factors limit the effectiveness of the computation, even in regimes where Type I ghost factors are suppressed. We introduced discernability as a measure of these limitations, which with the measurement uncertainty, sets an upper limit to the number of pulses that can be used. It also informs us about the cutoff choice, which sets the threshold that discriminates factors against nonfactors.

Furthermore, we introduced the use of preprocessing as a strategy to improve the discernability between factors and nonfactors, and demonstrated its use by experimentally implementing the scheme in a transmon qubit. Our experimental results corresponded well with the theory. Importantly, we showed that preprocessing enabled the Gauss sum factorization scheme to be useful even when the imperfect execution of the scheme in a NISQ device of the unmodified scheme gave the wrong results.

In the case where the pulse delay is much shorter than the pulse time, we required a phenomenological adjustment of the decoherence times as we discovered that the qubit decoheres at a different rate for different trial factors. We offered a likely explanation for this differing decoherence time with the filter function approach by demonstrating that the control pulses of a factor filters out a narrower range of the noise spectrum than that of a nonfactor, and that the nonfactor decohered faster than the factor for two common noise statistics.

However, we noted that the exact noise spectrum would be needed to confirm this quantitatively, and identified that a possible avenue of future research would be to perform noise spectroscopy on the environment to reconstruct the noise spectrum. This can lead to a better understanding of the system behavior when $\tau \ll t_\pi$, which would allow us to fit more operations within the coherence time of the qubit, and hence increase the upper limit to the number of digits that can be factorized.

\section{Data Availability}
The data that support the findings of this study are available from the corresponding authors on reasonable request.

\section{Acknowledgements}
LHZ was supported by the SGUnited program (CP0002392). YPT was supported by the SGUnited program (R00003494). We thank Dr Farshad Foroughi from Institut Néel for his assistance in the fabrication of the qubit sample used in the experiment.

\section{Additional Information}
The authors declare no conflict of interest.

\bibliographystyle{apsrev4-2}
%


\begin{thebibliography}{42}%
\makeatletter
\providecommand \@ifxundefined [1]{%
 \@ifx{#1\undefined}
}%
\providecommand \@ifnum [1]{%
 \ifnum #1\expandafter \@firstoftwo
 \else \expandafter \@secondoftwo
 \fi
}%
\providecommand \@ifx [1]{%
 \ifx #1\expandafter \@firstoftwo
 \else \expandafter \@secondoftwo
 \fi
}%
\providecommand \natexlab [1]{#1}%
\providecommand \enquote  [1]{``#1''}%
\providecommand \bibnamefont  [1]{#1}%
\providecommand \bibfnamefont [1]{#1}%
\providecommand \citenamefont [1]{#1}%
\providecommand \href@noop [0]{\@secondoftwo}%
\providecommand \href [0]{\begingroup \@sanitize@url \@href}%
\providecommand \@href[1]{\@@startlink{#1}\@@href}%
\providecommand \@@href[1]{\endgroup#1\@@endlink}%
\providecommand \@sanitize@url [0]{\catcode `\\12\catcode `\$12\catcode
  `\&12\catcode `\#12\catcode `\^12\catcode `\_12\catcode `\%12\relax}%
\providecommand \@@startlink[1]{}%
\providecommand \@@endlink[0]{}%
\providecommand \url  [0]{\begingroup\@sanitize@url \@url }%
\providecommand \@url [1]{\endgroup\@href {#1}{\urlprefix }}%
\providecommand \urlprefix  [0]{URL }%
\providecommand \Eprint [0]{\href }%
\providecommand \doibase [0]{https://doi.org/}%
\providecommand \selectlanguage [0]{\@gobble}%
\providecommand \bibinfo  [0]{\@secondoftwo}%
\providecommand \bibfield  [0]{\@secondoftwo}%
\providecommand \translation [1]{[#1]}%
\providecommand \BibitemOpen [0]{}%
\providecommand \bibitemStop [0]{}%
\providecommand \bibitemNoStop [0]{.\EOS\space}%
\providecommand \EOS [0]{\spacefactor3000\relax}%
\providecommand \BibitemShut  [1]{\csname bibitem#1\endcsname}%
\let\auto@bib@innerbib\@empty
\bibitem [{\citenamefont {{Cleve}}\ \emph {et~al.}(1998)\citenamefont
  {{Cleve}}, \citenamefont {{Ekert}}, \citenamefont {{Macchiavello}},\ and\
  \citenamefont {{Mosca}}}]{Cleve1998}%
  \BibitemOpen
  \bibfield  {author} {\bibinfo {author} {\bibfnamefont {R.}~\bibnamefont
  {{Cleve}}}, \bibinfo {author} {\bibfnamefont {A.}~\bibnamefont {{Ekert}}},
  \bibinfo {author} {\bibfnamefont {C.}~\bibnamefont {{Macchiavello}}},\ and\
  \bibinfo {author} {\bibfnamefont {M.}~\bibnamefont {{Mosca}}},\ }\href
  {https://doi.org/10.1098/rspa.1998.0164} {\bibfield  {journal} {\bibinfo
  {journal} {Proc. R. Soc. Lond. Ser. A}\ }\textbf {\bibinfo {volume} {454}},\
  \bibinfo {pages} {339} (\bibinfo {year} {1998})},\ \Eprint
  {https://arxiv.org/abs/quant-ph/9708016} {arXiv:quant-ph/9708016 [quant-ph]}
  \BibitemShut {NoStop}%
\bibitem [{\citenamefont {{Bennett}}\ and\ \citenamefont
  {{DiVincenzo}}(2000)}]{Bennett2000}%
  \BibitemOpen
  \bibfield  {author} {\bibinfo {author} {\bibfnamefont {C.~H.}\ \bibnamefont
  {{Bennett}}}\ and\ \bibinfo {author} {\bibfnamefont {D.~P.}\ \bibnamefont
  {{DiVincenzo}}},\ }\href {https://doi.org/10.1038/35005001} {\bibfield
  {journal} {\bibinfo  {journal} {\nat}\ }\textbf {\bibinfo {volume} {404}},\
  \bibinfo {pages} {247} (\bibinfo {year} {2000})}\BibitemShut {NoStop}%
\bibitem [{\citenamefont {{Lloyd}}(1999)}]{Lloyd2000}%
  \BibitemOpen
  \bibfield  {author} {\bibinfo {author} {\bibfnamefont {S.}~\bibnamefont
  {{Lloyd}}},\ }\href {https://doi.org/10.1103/PhysRevA.61.010301} {\bibfield
  {journal} {\bibinfo  {journal} {\pra}\ }\textbf {\bibinfo {volume} {61}},\
  \bibinfo {eid} {010301} (\bibinfo {year} {1999})},\ \Eprint
  {https://arxiv.org/abs/quant-ph/9903057} {arXiv:quant-ph/9903057 [quant-ph]}
  \BibitemShut {NoStop}%
\bibitem [{\citenamefont {{Stahlke}}(2014)}]{Stahlke2014}%
  \BibitemOpen
  \bibfield  {author} {\bibinfo {author} {\bibfnamefont {D.}~\bibnamefont
  {{Stahlke}}},\ }\href {https://doi.org/10.1103/PhysRevA.90.022302} {\bibfield
   {journal} {\bibinfo  {journal} {\pra}\ }\textbf {\bibinfo {volume} {90}},\
  \bibinfo {eid} {022302} (\bibinfo {year} {2014})},\ \Eprint
  {https://arxiv.org/abs/1305.2186} {arXiv:1305.2186 [quant-ph]} \BibitemShut
  {NoStop}%
\bibitem [{\citenamefont {Preskill}(2018)}]{Preskill2018}%
  \BibitemOpen
  \bibfield  {author} {\bibinfo {author} {\bibfnamefont {J.}~\bibnamefont
  {Preskill}},\ }\href {https://doi.org/10.22331/q-2018-08-06-79} {\bibfield
  {journal} {\bibinfo  {journal} {{Quantum}}\ }\textbf {\bibinfo {volume}
  {2}},\ \bibinfo {pages} {79} (\bibinfo {year} {2018})},\ \Eprint
  {https://arxiv.org/abs/1801.00862} {arXiv:1801.00862 [quant-ph]} \BibitemShut
  {NoStop}%
\bibitem [{\citenamefont {Shor}(1999)}]{Shor1999}%
  \BibitemOpen
  \bibfield  {author} {\bibinfo {author} {\bibfnamefont {P.~W.}\ \bibnamefont
  {Shor}},\ }\href {https://doi.org/10.1137/S0036144598347011} {\bibfield
  {journal} {\bibinfo  {journal} {SIAM Review}\ }\textbf {\bibinfo {volume}
  {41}},\ \bibinfo {pages} {303} (\bibinfo {year} {1999})},\ \Eprint
  {https://arxiv.org/abs/quant-ph/9508027} {arXiv:quant-ph/9508027 [quant-ph]}
  \BibitemShut {NoStop}%
\bibitem [{\citenamefont {Beauregard}(2003)}]{Beauregard2003}%
  \BibitemOpen
  \bibfield  {author} {\bibinfo {author} {\bibfnamefont {S.}~\bibnamefont
  {Beauregard}},\ }\href@noop {} {\bibfield  {journal} {\bibinfo  {journal}
  {Quantum Info. Comput.}\ }\textbf {\bibinfo {volume} {3}},\ \bibinfo {pages}
  {175–185} (\bibinfo {year} {2003})},\ \Eprint
  {https://arxiv.org/abs/quant-ph/0205095} {arXiv:quant-ph/0205095 [quant-ph]}
  \BibitemShut {NoStop}%
\bibitem [{\citenamefont {H{\"{a}}ner}\ \emph {et~al.}(2017)\citenamefont
  {H{\"{a}}ner}, \citenamefont {Roetteler},\ and\ \citenamefont
  {Svore}}]{Haner2017}%
  \BibitemOpen
  \bibfield  {author} {\bibinfo {author} {\bibfnamefont {T.}~\bibnamefont
  {H{\"{a}}ner}}, \bibinfo {author} {\bibfnamefont {M.}~\bibnamefont
  {Roetteler}},\ and\ \bibinfo {author} {\bibfnamefont {K.~M.}\ \bibnamefont
  {Svore}},\ }\href
  {http://www.rintonpress.com/xxqic17/qic-17-78/0673-0684.pdf} {\bibfield
  {journal} {\bibinfo  {journal} {Quantum Inf. Comput.}\ }\textbf {\bibinfo
  {volume} {17}},\ \bibinfo {pages} {673} (\bibinfo {year} {2017})},\ \Eprint
  {https://arxiv.org/abs/1611.07995} {arXiv:1611.07995 [quant-ph]} \BibitemShut
  {NoStop}%
\bibitem [{\citenamefont {{Skosana}}\ and\ \citenamefont
  {{Tame}}(2021)}]{Skosana2021}%
  \BibitemOpen
  \bibfield  {author} {\bibinfo {author} {\bibfnamefont {U.}~\bibnamefont
  {{Skosana}}}\ and\ \bibinfo {author} {\bibfnamefont {M.}~\bibnamefont
  {{Tame}}},\ }\href {https://doi.org/10.1038/s41598-021-95973-w} {\bibfield
  {journal} {\bibinfo  {journal} {Scientific Reports}\ }\textbf {\bibinfo
  {volume} {11}},\ \bibinfo {pages} {16599} (\bibinfo {year} {2021})},\ \Eprint
  {https://arxiv.org/abs/2103.13855} {arXiv:2103.13855 [quant-ph]} \BibitemShut
  {NoStop}%
\bibitem [{\citenamefont {Mart{\'i}n-L{\'o}pez}\ \emph
  {et~al.}(2012)\citenamefont {Mart{\'i}n-L{\'o}pez}, \citenamefont {Laing},
  \citenamefont {Lawson}, \citenamefont {Alvarez}, \citenamefont {Zhou},\ and\
  \citenamefont {O'Brien}}]{MartinLopez2012}%
  \BibitemOpen
  \bibfield  {author} {\bibinfo {author} {\bibfnamefont {E.}~\bibnamefont
  {Mart{\'i}n-L{\'o}pez}}, \bibinfo {author} {\bibfnamefont {A.}~\bibnamefont
  {Laing}}, \bibinfo {author} {\bibfnamefont {T.}~\bibnamefont {Lawson}},
  \bibinfo {author} {\bibfnamefont {R.}~\bibnamefont {Alvarez}}, \bibinfo
  {author} {\bibfnamefont {X.-Q.}\ \bibnamefont {Zhou}},\ and\ \bibinfo
  {author} {\bibfnamefont {J.~L.}\ \bibnamefont {O'Brien}},\ }\href
  {https://doi.org/10.1038/nphoton.2012.259} {\bibfield  {journal} {\bibinfo
  {journal} {Nature Photonics}\ }\textbf {\bibinfo {volume} {6}},\ \bibinfo
  {pages} {773} (\bibinfo {year} {2012})},\ \Eprint
  {https://arxiv.org/abs/1111.4147} {arXiv:1111.4147 [quant-ph]} \BibitemShut
  {NoStop}%
\bibitem [{\citenamefont {Devitt}\ \emph {et~al.}(2013)\citenamefont {Devitt},
  \citenamefont {Stephens}, \citenamefont {Munro},\ and\ \citenamefont
  {Nemoto}}]{Devitt2013}%
  \BibitemOpen
  \bibfield  {author} {\bibinfo {author} {\bibfnamefont {S.~J.}\ \bibnamefont
  {Devitt}}, \bibinfo {author} {\bibfnamefont {A.~M.}\ \bibnamefont
  {Stephens}}, \bibinfo {author} {\bibfnamefont {W.~J.}\ \bibnamefont
  {Munro}},\ and\ \bibinfo {author} {\bibfnamefont {K.}~\bibnamefont
  {Nemoto}},\ }\href {https://doi.org/10.1038/ncomms3524} {\bibfield  {journal}
  {\bibinfo  {journal} {Nature Communications}\ }\textbf {\bibinfo {volume}
  {4}},\ \bibinfo {pages} {2524} (\bibinfo {year} {2013})},\ \Eprint
  {https://arxiv.org/abs/1212.4934} {arXiv:1212.4934 [quant-ph]} \BibitemShut
  {NoStop}%
\bibitem [{\citenamefont {Peng}\ \emph {et~al.}(2008)\citenamefont {Peng},
  \citenamefont {Liao}, \citenamefont {Xu}, \citenamefont {Qin}, \citenamefont
  {Zhou}, \citenamefont {Suter},\ and\ \citenamefont {Du}}]{Peng2008}%
  \BibitemOpen
  \bibfield  {author} {\bibinfo {author} {\bibfnamefont {X.}~\bibnamefont
  {Peng}}, \bibinfo {author} {\bibfnamefont {Z.}~\bibnamefont {Liao}}, \bibinfo
  {author} {\bibfnamefont {N.}~\bibnamefont {Xu}}, \bibinfo {author}
  {\bibfnamefont {G.}~\bibnamefont {Qin}}, \bibinfo {author} {\bibfnamefont
  {X.}~\bibnamefont {Zhou}}, \bibinfo {author} {\bibfnamefont {D.}~\bibnamefont
  {Suter}},\ and\ \bibinfo {author} {\bibfnamefont {J.}~\bibnamefont {Du}},\
  }\href {https://doi.org/10.1103/PhysRevLett.101.220405} {\bibfield  {journal}
  {\bibinfo  {journal} {\prl}\ }\textbf {\bibinfo {volume} {101}},\ \bibinfo
  {pages} {220405} (\bibinfo {year} {2008})},\ \Eprint
  {https://arxiv.org/abs/0808.1935} {arXiv:0808.1935 [quant-ph]} \BibitemShut
  {NoStop}%
\bibitem [{\citenamefont {Dridi}\ and\ \citenamefont
  {Alghassi}(2017)}]{Dridi2017}%
  \BibitemOpen
  \bibfield  {author} {\bibinfo {author} {\bibfnamefont {R.}~\bibnamefont
  {Dridi}}\ and\ \bibinfo {author} {\bibfnamefont {H.}~\bibnamefont
  {Alghassi}},\ }\href {https://doi.org/10.1038/srep43048} {\bibfield
  {journal} {\bibinfo  {journal} {Scientific Reports}\ }\textbf {\bibinfo
  {volume} {7}},\ \bibinfo {pages} {43048} (\bibinfo {year} {2017})},\ \Eprint
  {https://arxiv.org/abs/1604.05796} {arXiv:1604.05796 [quant-ph]} \BibitemShut
  {NoStop}%
\bibitem [{\citenamefont {Jiang}\ \emph {et~al.}(2018)\citenamefont {Jiang},
  \citenamefont {Britt}, \citenamefont {McCaskey}, \citenamefont {Humble},\
  and\ \citenamefont {Kais}}]{Jiang2018}%
  \BibitemOpen
  \bibfield  {author} {\bibinfo {author} {\bibfnamefont {S.}~\bibnamefont
  {Jiang}}, \bibinfo {author} {\bibfnamefont {K.~A.}\ \bibnamefont {Britt}},
  \bibinfo {author} {\bibfnamefont {A.~J.}\ \bibnamefont {McCaskey}}, \bibinfo
  {author} {\bibfnamefont {T.~S.}\ \bibnamefont {Humble}},\ and\ \bibinfo
  {author} {\bibfnamefont {S.}~\bibnamefont {Kais}},\ }\href
  {https://doi.org/10.1038/s41598-018-36058-z} {\bibfield  {journal} {\bibinfo
  {journal} {Scientific Reports}\ }\textbf {\bibinfo {volume} {8}},\ \bibinfo
  {pages} {17667} (\bibinfo {year} {2018})},\ \Eprint
  {https://arxiv.org/abs/1804.02733} {arXiv:1804.02733 [quant-ph]} \BibitemShut
  {NoStop}%
\bibitem [{\citenamefont {Anschuetz}\ \emph {et~al.}(2019)\citenamefont
  {Anschuetz}, \citenamefont {Olson}, \citenamefont {Aspuru-Guzik},\ and\
  \citenamefont {Cao}}]{Anschuetz2018}%
  \BibitemOpen
  \bibfield  {author} {\bibinfo {author} {\bibfnamefont {E.}~\bibnamefont
  {Anschuetz}}, \bibinfo {author} {\bibfnamefont {J.}~\bibnamefont {Olson}},
  \bibinfo {author} {\bibfnamefont {A.}~\bibnamefont {Aspuru-Guzik}},\ and\
  \bibinfo {author} {\bibfnamefont {Y.}~\bibnamefont {Cao}},\ }in\ \href@noop
  {} {\emph {\bibinfo {booktitle} {Quantum Technology and Optimization
  Problems}}},\ \bibinfo {editor} {edited by\ \bibinfo {editor} {\bibfnamefont
  {S.}~\bibnamefont {Feld}}\ and\ \bibinfo {editor} {\bibfnamefont
  {C.}~\bibnamefont {Linnhoff-Popien}}}\ (\bibinfo  {publisher} {Springer
  International Publishing},\ \bibinfo {address} {Cham},\ \bibinfo {year}
  {2019})\ pp.\ \bibinfo {pages} {74--85},\ \Eprint
  {https://arxiv.org/abs/1808.08927} {arXiv:1808.08927 [quant-ph]} \BibitemShut
  {NoStop}%
\bibitem [{\citenamefont {Qiu}\ \emph {et~al.}(2020)\citenamefont {Qiu},
  \citenamefont {Alam}, \citenamefont {Ash-Saki},\ and\ \citenamefont
  {Ghosh}}]{Qiu2020}%
  \BibitemOpen
  \bibfield  {author} {\bibinfo {author} {\bibfnamefont {L.}~\bibnamefont
  {Qiu}}, \bibinfo {author} {\bibfnamefont {M.}~\bibnamefont {Alam}}, \bibinfo
  {author} {\bibfnamefont {A.}~\bibnamefont {Ash-Saki}},\ and\ \bibinfo
  {author} {\bibfnamefont {S.}~\bibnamefont {Ghosh}},\ }in\ \href
  {https://doi.org/10.1145/3370748.3406586} {\emph {\bibinfo {booktitle}
  {Proceedings of the ACM/IEEE International Symposium on Low Power Electronics
  and Design}}},\ \bibinfo {series and number} {ISLPED '20}\ (\bibinfo
  {publisher} {Association for Computing Machinery},\ \bibinfo {address} {New
  York, NY, USA},\ \bibinfo {year} {2020})\ p.\ \bibinfo {pages} {229–234},\
  \Eprint {https://arxiv.org/abs/2004.12447} {arXiv:2004.12447 [cs.ET]}
  \BibitemShut {NoStop}%
\bibitem [{\citenamefont {{Mack}}\ \emph {et~al.}(2002)\citenamefont {{Mack}},
  \citenamefont {{Bienert}}, \citenamefont {{Haug}}, \citenamefont
  {{Freyberger}},\ and\ \citenamefont {{Schleich}}}]{Mack2002}%
  \BibitemOpen
  \bibfield  {author} {\bibinfo {author} {\bibfnamefont {H.}~\bibnamefont
  {{Mack}}}, \bibinfo {author} {\bibfnamefont {M.}~\bibnamefont {{Bienert}}},
  \bibinfo {author} {\bibfnamefont {F.}~\bibnamefont {{Haug}}}, \bibinfo
  {author} {\bibfnamefont {M.}~\bibnamefont {{Freyberger}}},\ and\ \bibinfo
  {author} {\bibfnamefont {W.~P.}\ \bibnamefont {{Schleich}}},\ }\href
  {https://doi.org/10.1002/1521-3951(200210)233:3<408::AID-PSSB408>3.0.CO;2-N}
  {\bibfield  {journal} {\bibinfo  {journal} {Physica Status Solidi B Basic
  Research}\ }\textbf {\bibinfo {volume} {233}},\ \bibinfo {pages} {408}
  (\bibinfo {year} {2002})},\ \Eprint {https://arxiv.org/abs/quant-ph/0208021}
  {arXiv:quant-ph/0208021 [quant-ph]} \BibitemShut {NoStop}%
\bibitem [{\citenamefont {{Merkel}}\ \emph {et~al.}(2006)\citenamefont
  {{Merkel}}, \citenamefont {{Crasser}}, \citenamefont {{Haug}}, \citenamefont
  {{Lutz}}, \citenamefont {{Mack}}, \citenamefont {{Freyberger}}, \citenamefont
  {{Schleich}}, \citenamefont {{Averbukh}}, \citenamefont {{Bienert}},
  \citenamefont {{Girard}}, \citenamefont {{Maier}},\ and\ \citenamefont
  {{Paulus}}}]{Merkel2006}%
  \BibitemOpen
  \bibfield  {author} {\bibinfo {author} {\bibfnamefont {W.}~\bibnamefont
  {{Merkel}}}, \bibinfo {author} {\bibfnamefont {O.}~\bibnamefont {{Crasser}}},
  \bibinfo {author} {\bibfnamefont {F.}~\bibnamefont {{Haug}}}, \bibinfo
  {author} {\bibfnamefont {E.}~\bibnamefont {{Lutz}}}, \bibinfo {author}
  {\bibfnamefont {H.}~\bibnamefont {{Mack}}}, \bibinfo {author} {\bibfnamefont
  {M.}~\bibnamefont {{Freyberger}}}, \bibinfo {author} {\bibfnamefont {W.~P.}\
  \bibnamefont {{Schleich}}}, \bibinfo {author} {\bibfnamefont
  {I.}~\bibnamefont {{Averbukh}}}, \bibinfo {author} {\bibfnamefont
  {M.}~\bibnamefont {{Bienert}}}, \bibinfo {author} {\bibfnamefont
  {B.}~\bibnamefont {{Girard}}}, \bibinfo {author} {\bibfnamefont
  {H.}~\bibnamefont {{Maier}}},\ and\ \bibinfo {author} {\bibfnamefont {G.~G.}\
  \bibnamefont {{Paulus}}},\ }\href {https://doi.org/10.1142/S021797920603439X}
  {\bibfield  {journal} {\bibinfo  {journal} {International Journal of Modern
  Physics B}\ }\textbf {\bibinfo {volume} {20}},\ \bibinfo {pages} {1893}
  (\bibinfo {year} {2006})}\BibitemShut {NoStop}%
\bibitem [{\citenamefont {Rangelov}(2009)}]{Rangelov2009}%
  \BibitemOpen
  \bibfield  {author} {\bibinfo {author} {\bibfnamefont {A.~A.}\ \bibnamefont
  {Rangelov}},\ }\href {https://doi.org/10.1088/0953-4075/42/2/021002}
  {\bibfield  {journal} {\bibinfo  {journal} {Journal of Physics B: Atomic,
  Molecular and Optical Physics}\ }\textbf {\bibinfo {volume} {42}},\ \bibinfo
  {pages} {021002} (\bibinfo {year} {2009})},\ \Eprint
  {https://arxiv.org/abs/0811.2070} {arXiv:0811.2070 [quant-ph]} \BibitemShut
  {NoStop}%
\bibitem [{\citenamefont {{Mahesh}}\ \emph {et~al.}(2007)\citenamefont
  {{Mahesh}}, \citenamefont {{Rajendran}}, \citenamefont {{Peng}},\ and\
  \citenamefont {{Suter}}}]{Mahesh2007}%
  \BibitemOpen
  \bibfield  {author} {\bibinfo {author} {\bibfnamefont {T.~S.}\ \bibnamefont
  {{Mahesh}}}, \bibinfo {author} {\bibfnamefont {N.}~\bibnamefont
  {{Rajendran}}}, \bibinfo {author} {\bibfnamefont {X.}~\bibnamefont
  {{Peng}}},\ and\ \bibinfo {author} {\bibfnamefont {D.}~\bibnamefont
  {{Suter}}},\ }\href {https://doi.org/10.1103/PhysRevA.75.062303} {\bibfield
  {journal} {\bibinfo  {journal} {\pra}\ }\textbf {\bibinfo {volume} {75}},\
  \bibinfo {pages} {062303} (\bibinfo {year} {2007})},\ \Eprint
  {https://arxiv.org/abs/quant-ph/0701205} {arXiv:quant-ph/0701205 [quant-ph]}
  \BibitemShut {NoStop}%
\bibitem [{\citenamefont {{Mehring}}\ \emph {et~al.}(2007)\citenamefont
  {{Mehring}}, \citenamefont {{M{\"u}ller}}, \citenamefont {{Averbukh}},
  \citenamefont {{Merkel}},\ and\ \citenamefont {{Schleich}}}]{Mehring2007}%
  \BibitemOpen
  \bibfield  {author} {\bibinfo {author} {\bibfnamefont {M.}~\bibnamefont
  {{Mehring}}}, \bibinfo {author} {\bibfnamefont {K.}~\bibnamefont
  {{M{\"u}ller}}}, \bibinfo {author} {\bibfnamefont {I.~S.}\ \bibnamefont
  {{Averbukh}}}, \bibinfo {author} {\bibfnamefont {W.}~\bibnamefont
  {{Merkel}}},\ and\ \bibinfo {author} {\bibfnamefont {W.~P.}\ \bibnamefont
  {{Schleich}}},\ }\href {https://doi.org/10.1103/PhysRevLett.98.120502}
  {\bibfield  {journal} {\bibinfo  {journal} {\prl}\ }\textbf {\bibinfo
  {volume} {98}},\ \bibinfo {pages} {120502} (\bibinfo {year} {2007})},\
  \Eprint {https://arxiv.org/abs/quant-ph/0609174} {arXiv:quant-ph/0609174
  [quant-ph]} \BibitemShut {NoStop}%
\bibitem [{\citenamefont {Sadgrove}\ \emph {et~al.}(2008)\citenamefont
  {Sadgrove}, \citenamefont {Kumar},\ and\ \citenamefont
  {Nakagawa}}]{Sadgrove2008}%
  \BibitemOpen
  \bibfield  {author} {\bibinfo {author} {\bibfnamefont {M.}~\bibnamefont
  {Sadgrove}}, \bibinfo {author} {\bibfnamefont {S.}~\bibnamefont {Kumar}},\
  and\ \bibinfo {author} {\bibfnamefont {K.}~\bibnamefont {Nakagawa}},\ }\href
  {https://doi.org/10.1103/PhysRevLett.101.180502} {\bibfield  {journal}
  {\bibinfo  {journal} {Phys. Rev. Lett.}\ }\textbf {\bibinfo {volume} {101}},\
  \bibinfo {pages} {180502} (\bibinfo {year} {2008})}\BibitemShut {NoStop}%
\bibitem [{\citenamefont {Tamma}\ \emph {et~al.}(2011)\citenamefont {Tamma},
  \citenamefont {Zhang}, \citenamefont {He}, \citenamefont {Garuccio},
  \citenamefont {Schleich},\ and\ \citenamefont {Shih}}]{Tamma2011}%
  \BibitemOpen
  \bibfield  {author} {\bibinfo {author} {\bibfnamefont {V.}~\bibnamefont
  {Tamma}}, \bibinfo {author} {\bibfnamefont {H.}~\bibnamefont {Zhang}},
  \bibinfo {author} {\bibfnamefont {X.}~\bibnamefont {He}}, \bibinfo {author}
  {\bibfnamefont {A.}~\bibnamefont {Garuccio}}, \bibinfo {author}
  {\bibfnamefont {W.~P.}\ \bibnamefont {Schleich}},\ and\ \bibinfo {author}
  {\bibfnamefont {Y.}~\bibnamefont {Shih}},\ }\href
  {https://doi.org/10.1103/PhysRevA.83.020304} {\bibfield  {journal} {\bibinfo
  {journal} {Phys. Rev. A}\ }\textbf {\bibinfo {volume} {83}},\ \bibinfo
  {pages} {020304} (\bibinfo {year} {2011})},\ \Eprint
  {https://arxiv.org/abs/1506.02907} {arXiv:1506.02907 [quant-ph]} \BibitemShut
  {NoStop}%
\bibitem [{\citenamefont {Bigourd}\ \emph {et~al.}(2008)\citenamefont
  {Bigourd}, \citenamefont {Chatel}, \citenamefont {Schleich},\ and\
  \citenamefont {Girard}}]{Bigourd2008}%
  \BibitemOpen
  \bibfield  {author} {\bibinfo {author} {\bibfnamefont {D.}~\bibnamefont
  {Bigourd}}, \bibinfo {author} {\bibfnamefont {B.}~\bibnamefont {Chatel}},
  \bibinfo {author} {\bibfnamefont {W.~P.}\ \bibnamefont {Schleich}},\ and\
  \bibinfo {author} {\bibfnamefont {B.}~\bibnamefont {Girard}},\ }\href
  {https://doi.org/10.1103/PhysRevLett.100.030202} {\bibfield  {journal}
  {\bibinfo  {journal} {Phys. Rev. Lett.}\ }\textbf {\bibinfo {volume} {100}},\
  \bibinfo {pages} {030202} (\bibinfo {year} {2008})},\ \Eprint
  {https://arxiv.org/abs/0709.1906} {arXiv:0709.1906 [physics.optics]}
  \BibitemShut {NoStop}%
\bibitem [{\citenamefont {Pelka}\ \emph {et~al.}(2018)\citenamefont {Pelka},
  \citenamefont {Graf}, \citenamefont {Mehringer},\ and\ \citenamefont {von
  Zanthier}}]{Pelka2018}%
  \BibitemOpen
  \bibfield  {author} {\bibinfo {author} {\bibfnamefont {K.}~\bibnamefont
  {Pelka}}, \bibinfo {author} {\bibfnamefont {J.}~\bibnamefont {Graf}},
  \bibinfo {author} {\bibfnamefont {T.}~\bibnamefont {Mehringer}},\ and\
  \bibinfo {author} {\bibfnamefont {J.}~\bibnamefont {von Zanthier}},\ }\href
  {https://doi.org/10.1364/OE.26.015009} {\bibfield  {journal} {\bibinfo
  {journal} {Opt. Express}\ }\textbf {\bibinfo {volume} {26}},\ \bibinfo
  {pages} {15009} (\bibinfo {year} {2018})},\ \Eprint
  {https://arxiv.org/abs/1803.01559} {arXiv:1803.01559 [physics.optics]}
  \BibitemShut {NoStop}%
\bibitem [{\citenamefont {{Ng}}\ and\ \citenamefont {{Nori}}(2010)}]{Ng2010}%
  \BibitemOpen
  \bibfield  {author} {\bibinfo {author} {\bibfnamefont {H.~T.}\ \bibnamefont
  {{Ng}}}\ and\ \bibinfo {author} {\bibfnamefont {F.}~\bibnamefont {{Nori}}},\
  }\href {https://doi.org/10.1103/PhysRevA.82.042317} {\bibfield  {journal}
  {\bibinfo  {journal} {\pra}\ }\textbf {\bibinfo {volume} {82}},\ \bibinfo
  {pages} {042317} (\bibinfo {year} {2010})},\ \Eprint
  {https://arxiv.org/abs/0911.0249} {arXiv:0911.0249 [quant-ph]} \BibitemShut
  {NoStop}%
\bibitem [{\citenamefont {Merkel}\ \emph {et~al.}(2011)\citenamefont {Merkel},
  \citenamefont {Wölk}, \citenamefont {Schleich}, \citenamefont {Averbukh},
  \citenamefont {Girard},\ and\ \citenamefont {Paulus}}]{Merkel2011}%
  \BibitemOpen
  \bibfield  {author} {\bibinfo {author} {\bibfnamefont {W.}~\bibnamefont
  {Merkel}}, \bibinfo {author} {\bibfnamefont {S.}~\bibnamefont {Wölk}},
  \bibinfo {author} {\bibfnamefont {W.~P.}\ \bibnamefont {Schleich}}, \bibinfo
  {author} {\bibfnamefont {I.~S.}\ \bibnamefont {Averbukh}}, \bibinfo {author}
  {\bibfnamefont {B.}~\bibnamefont {Girard}},\ and\ \bibinfo {author}
  {\bibfnamefont {G.~G.}\ \bibnamefont {Paulus}},\ }\href
  {https://doi.org/10.1088/1367-2630/13/10/103008} {\bibfield  {journal}
  {\bibinfo  {journal} {New Journal of Physics}\ }\textbf {\bibinfo {volume}
  {13}},\ \bibinfo {pages} {103008} (\bibinfo {year} {2011})},\ \Eprint
  {https://arxiv.org/abs/1210.6487} {arXiv:1210.6487 [quant-ph]} \BibitemShut
  {NoStop}%
\bibitem [{\citenamefont {{Li}}\ \emph {et~al.}(2012)\citenamefont {{Li}},
  \citenamefont {{Peng}}, \citenamefont {{Du}},\ and\ \citenamefont
  {{Suter}}}]{Jun2011}%
  \BibitemOpen
  \bibfield  {author} {\bibinfo {author} {\bibfnamefont {J.}~\bibnamefont
  {{Li}}}, \bibinfo {author} {\bibfnamefont {X.}~\bibnamefont {{Peng}}},
  \bibinfo {author} {\bibfnamefont {J.}~\bibnamefont {{Du}}},\ and\ \bibinfo
  {author} {\bibfnamefont {D.}~\bibnamefont {{Suter}}},\ }\href
  {https://doi.org/10.1038/srep00260} {\bibfield  {journal} {\bibinfo
  {journal} {Scientific Reports}\ }\textbf {\bibinfo {volume} {2}},\ \bibinfo
  {pages} {260} (\bibinfo {year} {2012})},\ \Eprint
  {https://arxiv.org/abs/1108.5848} {arXiv:1108.5848 [quant-ph]} \BibitemShut
  {NoStop}%
\bibitem [{\citenamefont {Wölk}\ and\ \citenamefont
  {Schleich}(2012)}]{Wolk2012}%
  \BibitemOpen
  \bibfield  {author} {\bibinfo {author} {\bibfnamefont {S.}~\bibnamefont
  {Wölk}}\ and\ \bibinfo {author} {\bibfnamefont {W.~P.}\ \bibnamefont
  {Schleich}},\ }\href {https://doi.org/10.1088/1367-2630/14/1/013049}
  {\bibfield  {journal} {\bibinfo  {journal} {New Journal of Physics}\ }\textbf
  {\bibinfo {volume} {14}},\ \bibinfo {pages} {013049} (\bibinfo {year}
  {2012})},\ \Eprint {https://arxiv.org/abs/1210.6491} {arXiv:1210.6491
  [quant-ph]} \BibitemShut {NoStop}%
\bibitem [{\citenamefont {Wölk}\ \emph {et~al.}(2011)\citenamefont {Wölk},
  \citenamefont {Merkel}, \citenamefont {Schleich}, \citenamefont {Averbukh},\
  and\ \citenamefont {Girard}}]{Wolk2011}%
  \BibitemOpen
  \bibfield  {author} {\bibinfo {author} {\bibfnamefont {S.}~\bibnamefont
  {Wölk}}, \bibinfo {author} {\bibfnamefont {W.}~\bibnamefont {Merkel}},
  \bibinfo {author} {\bibfnamefont {W.~P.}\ \bibnamefont {Schleich}}, \bibinfo
  {author} {\bibfnamefont {I.~S.}\ \bibnamefont {Averbukh}},\ and\ \bibinfo
  {author} {\bibfnamefont {B.}~\bibnamefont {Girard}},\ }\href
  {https://doi.org/10.1088/1367-2630/13/10/103007} {\bibfield  {journal}
  {\bibinfo  {journal} {New Journal of Physics}\ }\textbf {\bibinfo {volume}
  {13}},\ \bibinfo {pages} {103007} (\bibinfo {year} {2011})},\ \Eprint
  {https://arxiv.org/abs/1210.6474} {arXiv:1210.6474 [quant-ph]} \BibitemShut
  {NoStop}%
\bibitem [{\citenamefont {{\v{S}}tefa{\v{n}}{\'{a}}k}\ \emph
  {et~al.}(2007)\citenamefont {{\v{S}}tefa{\v{n}}{\'{a}}k}, \citenamefont
  {Merkel}, \citenamefont {Schleich}, \citenamefont {Haase},\ and\
  \citenamefont {Maier}}]{Stefanak2007}%
  \BibitemOpen
  \bibfield  {author} {\bibinfo {author} {\bibfnamefont {M.}~\bibnamefont
  {{\v{S}}tefa{\v{n}}{\'{a}}k}}, \bibinfo {author} {\bibfnamefont
  {W.}~\bibnamefont {Merkel}}, \bibinfo {author} {\bibfnamefont {W.~P.}\
  \bibnamefont {Schleich}}, \bibinfo {author} {\bibfnamefont {D.}~\bibnamefont
  {Haase}},\ and\ \bibinfo {author} {\bibfnamefont {H.}~\bibnamefont {Maier}},\
  }\href {https://doi.org/10.1088/1367-2630/9/10/370} {\bibfield  {journal}
  {\bibinfo  {journal} {New Journal of Physics}\ }\textbf {\bibinfo {volume}
  {9}},\ \bibinfo {pages} {370} (\bibinfo {year} {2007})}\BibitemShut {NoStop}%
\bibitem [{\citenamefont {Weber}\ \emph {et~al.}(2008)\citenamefont {Weber},
  \citenamefont {Chatel},\ and\ \citenamefont {Girard}}]{Weber2008}%
  \BibitemOpen
  \bibfield  {author} {\bibinfo {author} {\bibfnamefont {S.}~\bibnamefont
  {Weber}}, \bibinfo {author} {\bibfnamefont {B.}~\bibnamefont {Chatel}},\ and\
  \bibinfo {author} {\bibfnamefont {B.}~\bibnamefont {Girard}},\ }\href
  {https://doi.org/10.1209/0295-5075/83/34008} {\bibfield  {journal} {\bibinfo
  {journal} {{EPL} (Europhysics Letters)}\ }\textbf {\bibinfo {volume} {83}},\
  \bibinfo {pages} {34008} (\bibinfo {year} {2008})}\BibitemShut {NoStop}%
\bibitem [{\citenamefont {{Krantz}}\ \emph {et~al.}(2019)\citenamefont
  {{Krantz}}, \citenamefont {{Kjaergaard}}, \citenamefont {{Yan}},
  \citenamefont {{Orlando}}, \citenamefont {{Gustavsson}},\ and\ \citenamefont
  {{Oliver}}}]{Krantz2019}%
  \BibitemOpen
  \bibfield  {author} {\bibinfo {author} {\bibfnamefont {P.}~\bibnamefont
  {{Krantz}}}, \bibinfo {author} {\bibfnamefont {M.}~\bibnamefont
  {{Kjaergaard}}}, \bibinfo {author} {\bibfnamefont {F.}~\bibnamefont {{Yan}}},
  \bibinfo {author} {\bibfnamefont {T.~P.}\ \bibnamefont {{Orlando}}}, \bibinfo
  {author} {\bibfnamefont {S.}~\bibnamefont {{Gustavsson}}},\ and\ \bibinfo
  {author} {\bibfnamefont {W.~D.}\ \bibnamefont {{Oliver}}},\ }\href
  {https://doi.org/10.1063/1.5089550} {\bibfield  {journal} {\bibinfo
  {journal} {Applied Physics Reviews}\ }\textbf {\bibinfo {volume} {6}},\
  \bibinfo {eid} {021318} (\bibinfo {year} {2019})},\ \Eprint
  {https://arxiv.org/abs/1904.06560} {arXiv:1904.06560 [quant-ph]} \BibitemShut
  {NoStop}%
\bibitem [{\citenamefont {Berry}\ and\ \citenamefont
  {Goldberg}(1988)}]{Berry1988}%
  \BibitemOpen
  \bibfield  {author} {\bibinfo {author} {\bibfnamefont {M.~V.}\ \bibnamefont
  {Berry}}\ and\ \bibinfo {author} {\bibfnamefont {J.}~\bibnamefont
  {Goldberg}},\ }\href {https://doi.org/10.1088/0951-7715/1/1/001} {\bibfield
  {journal} {\bibinfo  {journal} {Nonlinearity}\ }\textbf {\bibinfo {volume}
  {1}},\ \bibinfo {pages} {1} (\bibinfo {year} {1988})}\BibitemShut {NoStop}%
\bibitem [{\citenamefont {Gauß}(1808)}]{Gauss1808}%
  \BibitemOpen
  \bibfield  {author} {\bibinfo {author} {\bibfnamefont {C.~F.}\ \bibnamefont
  {Gauß}},\ }\href@noop {} {\emph {\bibinfo {title} {Summatio quarumdam
  serierum singularium}}}\ (\bibinfo  {publisher} {Dieterich},\ \bibinfo
  {address} {Gottingae},\ \bibinfo {year} {1808})\BibitemShut {NoStop}%
\bibitem [{\citenamefont {Gilowski}\ \emph {et~al.}(2008)\citenamefont
  {Gilowski}, \citenamefont {Wendrich}, \citenamefont {M\"uller}, \citenamefont
  {Jentsch}, \citenamefont {Ertmer}, \citenamefont {Rasel},\ and\ \citenamefont
  {Schleich}}]{Gilowski2008}%
  \BibitemOpen
  \bibfield  {author} {\bibinfo {author} {\bibfnamefont {M.}~\bibnamefont
  {Gilowski}}, \bibinfo {author} {\bibfnamefont {T.}~\bibnamefont {Wendrich}},
  \bibinfo {author} {\bibfnamefont {T.}~\bibnamefont {M\"uller}}, \bibinfo
  {author} {\bibfnamefont {C.}~\bibnamefont {Jentsch}}, \bibinfo {author}
  {\bibfnamefont {W.}~\bibnamefont {Ertmer}}, \bibinfo {author} {\bibfnamefont
  {E.~M.}\ \bibnamefont {Rasel}},\ and\ \bibinfo {author} {\bibfnamefont
  {W.~P.}\ \bibnamefont {Schleich}},\ }\href
  {https://doi.org/10.1103/PhysRevLett.100.030201} {\bibfield  {journal}
  {\bibinfo  {journal} {Phys. Rev. Lett.}\ }\textbf {\bibinfo {volume} {100}},\
  \bibinfo {pages} {030201} (\bibinfo {year} {2008})},\ \Eprint
  {https://arxiv.org/abs/0709.1424} {arXiv:0709.1424 [quant-ph]} \BibitemShut
  {NoStop}%
\bibitem [{\citenamefont {Born}\ \emph {et~al.}(1999)\citenamefont {Born},
  \citenamefont {Wolf}, \citenamefont {Bhatia}, \citenamefont {Clemmow},
  \citenamefont {Gabor}, \citenamefont {Stokes}, \citenamefont {Taylor},
  \citenamefont {Wayman},\ and\ \citenamefont {Wilcock}}]{Born1999}%
  \BibitemOpen
  \bibfield  {author} {\bibinfo {author} {\bibfnamefont {M.}~\bibnamefont
  {Born}}, \bibinfo {author} {\bibfnamefont {E.}~\bibnamefont {Wolf}}, \bibinfo
  {author} {\bibfnamefont {A.~B.}\ \bibnamefont {Bhatia}}, \bibinfo {author}
  {\bibfnamefont {P.~C.}\ \bibnamefont {Clemmow}}, \bibinfo {author}
  {\bibfnamefont {D.}~\bibnamefont {Gabor}}, \bibinfo {author} {\bibfnamefont
  {A.~R.}\ \bibnamefont {Stokes}}, \bibinfo {author} {\bibfnamefont {A.~M.}\
  \bibnamefont {Taylor}}, \bibinfo {author} {\bibfnamefont {P.~A.}\
  \bibnamefont {Wayman}},\ and\ \bibinfo {author} {\bibfnamefont {W.~L.}\
  \bibnamefont {Wilcock}},\ }\href {https://doi.org/10.1017/CBO9781139644181}
  {\emph {\bibinfo {title} {Principles of Optics: Electromagnetic Theory of
  Propagation, Interference and Diffraction of Light}}},\ \bibinfo {edition}
  {7th}\ ed.\ (\bibinfo  {publisher} {Cambridge University Press},\ \bibinfo
  {year} {1999})\BibitemShut {NoStop}%
\bibitem [{\citenamefont {{Green}}\ \emph {et~al.}(2013)\citenamefont
  {{Green}}, \citenamefont {{Sastrawan}}, \citenamefont {{Uys}},\ and\
  \citenamefont {{Biercuk}}}]{Green2013}%
  \BibitemOpen
  \bibfield  {author} {\bibinfo {author} {\bibfnamefont {T.~J.}\ \bibnamefont
  {{Green}}}, \bibinfo {author} {\bibfnamefont {J.}~\bibnamefont
  {{Sastrawan}}}, \bibinfo {author} {\bibfnamefont {H.}~\bibnamefont {{Uys}}},\
  and\ \bibinfo {author} {\bibfnamefont {M.~J.}\ \bibnamefont {{Biercuk}}},\
  }\href {https://doi.org/10.1088/1367-2630/15/9/095004} {\bibfield  {journal}
  {\bibinfo  {journal} {New Journal of Physics}\ }\textbf {\bibinfo {volume}
  {15}},\ \bibinfo {eid} {095004} (\bibinfo {year} {2013})},\ \Eprint
  {https://arxiv.org/abs/1211.1163} {arXiv:1211.1163 [quant-ph]} \BibitemShut
  {NoStop}%
\bibitem [{\citenamefont {Bylander}\ \emph {et~al.}(2011)\citenamefont
  {Bylander}, \citenamefont {Gustavsson}, \citenamefont {Yan}, \citenamefont
  {Yoshihara}, \citenamefont {Harrabi}, \citenamefont {Fitch}, \citenamefont
  {Cory}, \citenamefont {Nakamura}, \citenamefont {Tsai},\ and\ \citenamefont
  {Oliver}}]{Bylander2011}%
  \BibitemOpen
  \bibfield  {author} {\bibinfo {author} {\bibfnamefont {J.}~\bibnamefont
  {Bylander}}, \bibinfo {author} {\bibfnamefont {S.}~\bibnamefont
  {Gustavsson}}, \bibinfo {author} {\bibfnamefont {F.}~\bibnamefont {Yan}},
  \bibinfo {author} {\bibfnamefont {F.}~\bibnamefont {Yoshihara}}, \bibinfo
  {author} {\bibfnamefont {K.}~\bibnamefont {Harrabi}}, \bibinfo {author}
  {\bibfnamefont {G.}~\bibnamefont {Fitch}}, \bibinfo {author} {\bibfnamefont
  {D.~G.}\ \bibnamefont {Cory}}, \bibinfo {author} {\bibfnamefont
  {Y.}~\bibnamefont {Nakamura}}, \bibinfo {author} {\bibfnamefont {J.-S.}\
  \bibnamefont {Tsai}},\ and\ \bibinfo {author} {\bibfnamefont {W.~D.}\
  \bibnamefont {Oliver}},\ }\href {https://doi.org/10.1038/nphys1994}
  {\bibfield  {journal} {\bibinfo  {journal} {Nature Physics}\ }\textbf
  {\bibinfo {volume} {7}},\ \bibinfo {pages} {565} (\bibinfo {year} {2011})},\
  \Eprint {https://arxiv.org/abs/1101.4707} {arXiv:1101.4707
  [cond-mat.supr-con]} \BibitemShut {NoStop}%
\bibitem [{\citenamefont {Paladino}\ \emph {et~al.}(2002)\citenamefont
  {Paladino}, \citenamefont {Faoro}, \citenamefont {Falci},\ and\ \citenamefont
  {Fazio}}]{Paladino2002}%
  \BibitemOpen
  \bibfield  {author} {\bibinfo {author} {\bibfnamefont {E.}~\bibnamefont
  {Paladino}}, \bibinfo {author} {\bibfnamefont {L.}~\bibnamefont {Faoro}},
  \bibinfo {author} {\bibfnamefont {G.}~\bibnamefont {Falci}},\ and\ \bibinfo
  {author} {\bibfnamefont {R.}~\bibnamefont {Fazio}},\ }\href
  {https://doi.org/10.1103/PhysRevLett.88.228304} {\bibfield  {journal}
  {\bibinfo  {journal} {Phys. Rev. Lett.}\ }\textbf {\bibinfo {volume} {88}},\
  \bibinfo {pages} {228304} (\bibinfo {year} {2002})},\ \Eprint
  {https://arxiv.org/abs/cond-mat/0201449} {arXiv:cond-mat/0201449
  [cond-mat.mes-hall]} \BibitemShut {NoStop}%
\bibitem [{\citenamefont {Koch}\ \emph {et~al.}(2007)\citenamefont {Koch},
  \citenamefont {DiVincenzo},\ and\ \citenamefont {Clarke}}]{Koch2007}%
  \BibitemOpen
  \bibfield  {author} {\bibinfo {author} {\bibfnamefont {R.~H.}\ \bibnamefont
  {Koch}}, \bibinfo {author} {\bibfnamefont {D.~P.}\ \bibnamefont
  {DiVincenzo}},\ and\ \bibinfo {author} {\bibfnamefont {J.}~\bibnamefont
  {Clarke}},\ }\href {https://doi.org/10.1103/PhysRevLett.98.267003} {\bibfield
   {journal} {\bibinfo  {journal} {Phys. Rev. Lett.}\ }\textbf {\bibinfo
  {volume} {98}},\ \bibinfo {pages} {267003} (\bibinfo {year}
  {2007})}\BibitemShut {NoStop}%
\bibitem [{\citenamefont {Altmann}(1989)}]{Altmann1989}%
  \BibitemOpen
  \bibfield  {author} {\bibinfo {author} {\bibfnamefont {S.~L.}\ \bibnamefont
  {Altmann}},\ }\href {https://doi.org/10.1080/0025570X.1989.11977459}
  {\bibfield  {journal} {\bibinfo  {journal} {Mathematics Magazine}\ }\textbf
  {\bibinfo {volume} {62}},\ \bibinfo {pages} {291} (\bibinfo {year} {1989})},\
  \Eprint
  {https://arxiv.org/abs/https://doi.org/10.1080/0025570X.1989.11977459}
  {https://doi.org/10.1080/0025570X.1989.11977459} \BibitemShut {NoStop}%
\end{thebibliography}

\appendix
\section{Resultant Rotations\label{apd:resultant-rotations}}
With the pulse sequence shown in Fig.~\ref{fig:circuit}, the overall operation on the qubit is the resultant rotation $\hat{R}(\gamma_m,\hat{n}_m) = \prod_{k=0}^{k=m}\hat{R}(\pi,\phi_k)$. We find the final rotation using the Rodriguez formula for the product of two rotations \cite{Altmann1989} $\hat{R}(\gamma_m,\hat{n}_m) = \hat{R}(\pi,\phi_m)\hat{R}(\gamma_{m-1},\hat{n}_{m-1})$ with
\begin{align}
\cos\pqty{\frac{\gamma_m}{2}} &= \cos\pqty{\frac{\gamma_{m-1}}{2}}\cos\pqty{\frac{\pi}{2}} \\
&\qquad -\sin\pqty{\frac{\gamma_{m-1}}{2}}\sin\pqty{\frac{\pi}{2}} \hat{n}_{\phi_m}\cdot\hat{n}_{m-1} \nonumber\\
&= -\sin\pqty{\frac{\gamma_{m-1}}{2}}\hat{n}_{\phi_m}\cdot\hat{n}_{m-1}\nonumber\\
\sin\pqty{\frac{\gamma_m}{2}}\hat{n}_m &=  \sin\pqty{\frac{\gamma_{m-1}}{2}}\cos\pqty{\frac{\pi}{2}}\hat{n}_{m-1} \nonumber\\
&\qquad+ \cos\pqty{\frac{\gamma_{m-1}}{2}}\sin\pqty{\frac{\pi}{2}}\hat{n}_{\phi_m}\\
&\qquad+ \sin\pqty{\frac{\gamma_{m-1}}{2}}\sin\pqty{\frac{\pi}{2}}\hat{n}_{\phi_m}\times\hat{n}_{m-1} \nonumber\\
&=\cos\pqty{\frac{\gamma_{m-1}}{2}}\hat{n}_{\phi_m}
+ \sin\pqty{\frac{\gamma_{m-1}}{2}}\hat{n}_{\phi_m}\times\hat{n}_{m-1}\nonumber
\end{align}
where $\hat{n}_{\phi_m} = \cos(\phi_m)\hat{x} + \sin(\phi_m)\hat{y}$. This works out to be
\begin{equation}
\hat{R}(\gamma_{m},\hat{n}_{m}) = \begin{cases}\cos(\pi m^2 \frac{N}{l})\hat{\mathbbm{1}} + i\sin(\pi m^2 \frac{N}{l})\hat{\sigma}_z & m\text{ even}, \\
\cos(\pi m^2 \frac{N}{l})\hat{\sigma}_x + \sin(\pi m^2 \frac{N}{l})\hat{\sigma}_y & m\text{ odd}.
\end{cases}
\end{equation}
In our setup, we have used $\phi_i = \phi_f = \pi/2$, and measure the probability of finding the system in the $\ket{1}$ state,
\begin{equation}
\begin{aligned}
\Pr(m;l,N) &= \abs{\mel{0}{\hat{R}\pqty{\tfrac{\pi}{2},\hat{y}}\hat{R}(\gamma_m,\hat{n}_m)\hat{R}\pqty{\tfrac{\pi}{2},\hat{y}}}{1}}^2 \\
&=
\abs{\ev{\hat{R}(\gamma_m,\hat{n}_m)}{+}}^2\\
&= \pqty{\cos(\pi m^2 \frac{N}{l})}^2\\
&= \frac{1}{2}\pqty{
1 + \cos(2\pi m^2 \frac{N}{l})
}.
\end{aligned}
\end{equation}
\section{Gauss Sum Filter Function\label{apd:gauss-sum-filter-function}}
The filter function is defined as
\begin{align}
g_{ij}(\omega) &= \frac{1}{\omega^2}\bqty{ R_\omega(\omega)R_\omega^\dagger(\omega)}_{ij},\tag{\ref{eq:filter-function} in main text}\\
R_\omega(\omega) &= -i\omega\int_0^t\dd{t'} R^{(\text{ctrl})}(t') \, e^{i\omega t'}\tag{\ref{eq:rotation-frequency-domain} in main text}.
\end{align}
For the Gauss sum sequence we have defined, for a particular value of $m$, given that the qubit rotates at a frequency $\Omega$ and takes $t_\pi = \pi/\Omega$ time for a $\pi$-pulse,
\begin{widetext}
\begin{align}
&R_\omega(\omega;m) \nonumber\\
&=
-i\omega\int_0^t\dd{t'}\;R^{(\text{ctrl})}(t')e^{i\omega t'} \nonumber\\
&=
-i\omega\Bigg\{ \nonumber\\
	&\qquad\qquad \mathbbm{1} \bigg[\int_0^{\frac{\tau}{2}}\dd{t'}\; e^{i\omega t'}  +
	\int_{\frac{\tau}{2}}^{\frac{\tau}{2}+t_\pi}\dd{t'}\;R\pqty{\Omega (t'-\tfrac{\tau}{2}),\phi_0} e^{i\omega t'}  +
	R(\pi,\phi_0)\int_{\frac{\tau}{2}+t_\pi}^{\tau+t_\pi}\dd{t'} e^{i\omega t'}\bigg] + \nonumber\\
	&\qquad\qquad R(\pi,\phi_0)\bigg[
		\int_{(\tau+t_\pi)}^{(\tau+t_\pi)+\frac{\tau}{2}}\dd{t'}\; e^{i\omega t'} + 
	    \int_{(\tau+t_\pi)+\frac{\tau}{2}}^{(\tau+t_\pi)+\frac{\tau}{2}+t_\pi}\dd{t'}\;R\pqty{\Omega (t'-((\tau+t_\pi)+\tfrac{\tau}{2})),\phi_1} e^{i\omega t'} + \nonumber\\
	 &\qquad\qquad\qquad\qquad R(\pi,\phi_1)\int_{(\tau+t_\pi)+\frac{\tau}{2}+t_\pi}^{(\tau+t_\pi)+\tau+t_\pi}\dd{t'} e^{i\omega t'} \bigg]+ \nonumber\\
 	&\qquad\qquad R(\pi,\phi_0) R(\pi,\phi_1)\bigg[
 		\int_{2(\tau+t_\pi)}^{2(\tau+t_\pi)+\frac{\tau}{2}}\dd{t'}\; e^{i\omega t'} + 
 		\int_{2(\tau+t_\pi)+\frac{\tau}{2}}^{2(\tau+t_\pi)+\frac{\tau}{2}+t_\pi}\dd{t'}\;R\pqty{\Omega (t'-(2(\tau+t_\pi)+\tfrac{\tau}{2})),\phi_2} e^{i\omega t'} + \nonumber\\
 	 &\qquad\qquad\qquad\qquad\qquad\qquad R(\pi,\phi_2)\int_{2(\tau+t_\pi)+\frac{\tau}{2}+t_\pi}^{2(\tau+t_\pi)+\tau+t_\pi}\dd{t'} e^{i\omega t'} \bigg] + \nonumber\\
&\qquad\qquad \cdots \nonumber\\
&\qquad\quad\Bigg\}\\
&= -i\omega \Bigg\{ \nonumber\\
	&\qquad\qquad \mathbbm{1} \bigg[\frac{e^{i\omega\frac{\tau}{2}}-1}{i\omega} +
	e^{i\omega\frac{\tau}{2}}\int_{0}^{t_\pi}\dd{t'}\;R\pqty{\Omega t',\phi_0} e^{i\omega t'} + 
	e^{i\omega(\frac{\tau}{2}+t_\pi)}\frac{e^{i\omega\frac{\tau}{2}}-1}{i\omega}R(\pi,\phi_0)\bigg] + \nonumber\\
		&\qquad\qquad R(\gamma_0,\hat{n}_0) e^{i\omega(\tau+t_\pi)}\bigg[\frac{e^{i\omega\frac{\tau}{2}}-1}{i\omega} +
		e^{i\omega\frac{\tau}{2}}\int_{0}^{t_\pi}\dd{t'}\;R\pqty{\Omega t',\phi_1} e^{i\omega t'} +
		e^{i\omega(\frac{\tau}{2}+t_\pi)}\frac{e^{i\omega\frac{\tau}{2}}-1}{i\omega}R(\pi,\phi_1)\bigg] + \nonumber\\
		&\qquad\qquad R(\gamma_1,\hat{n}_1) e^{i\omega 2(\tau+t_\pi)}\bigg[\frac{e^{i\omega\frac{\tau}{2}}-1}{i\omega} +
			e^{i\omega\frac{\tau}{2}}\int_{0}^{t_\pi}\dd{t'}\;R\pqty{\Omega t',\phi_2} e^{i\omega t'} +
			e^{i\omega(\frac{\tau}{2}+t_\pi)}\frac{e^{i\omega\frac{\tau}{2}}-1}{i\omega}R(\pi,\phi_2)\bigg]+ \nonumber\\
&\qquad\qquad\cdots \nonumber\\
&\qquad\quad\Bigg\} \nonumber\\
&= \sum_{k=0}^m R(\gamma_{k-1},\hat{n}_{k-1})e^{ik\omega(\tau+t_\pi)}\bigg[
	\pqty{1-e^{i\omega\frac{\tau}{2}}}\pqty{\mathbbm{1} + e^{i\omega(\frac{\tau}{2}+t_\pi)} R(\pi,\phi_k)} -
	i\omega e^{i\omega\frac{\tau}{2}}\int_{0}^{t_\pi}\dd{t'}\;R\pqty{\Omega t',\phi_k} e^{i\omega t'}
\bigg]\nonumber
\end{align}
Here, $R(\gamma_{-1},\hat{n}_{-1}) \equiv \mathbbm{1}$ for the $m=0$ case to hold. From Eq.~\ref{eq:gauss-angles}, $\phi_0 = 0$ for $m=0$ independent of the trial factor,
\begin{equation}
R_\omega(m=0)
= \pqty{1-e^{-\omega\frac{\tau}{2}}}{\scriptstyle\begin{pmatrix}
		1 + e^{i\omega\pqty{\frac{\tau}{2}+t_\pi}} & 0 & 0 \\
		0 & 1 - e^{i\omega\pqty{\frac{\tau}{2}+t_\pi}} & 0 \\
		0 & 0 & 1 - e^{i\omega\pqty{\frac{\tau}{2}+t_\pi}} \\
	\end{pmatrix}} + e^{-\omega\frac{\tau}{2}}\frac{1+e^{i\omega t_\pi}}{\omega^2-\Omega^2}{\scriptstyle\begin{pmatrix}
		\frac{\omega^2-\Omega^2}{1+e^{i\omega t_\pi}}\pqty{1-e^{i\omega t_\pi}} & 0 & 0 \\
		0 & \omega^2 & -i\omega\Omega  \\
		0 & i\omega\Omega & \omega^2  \\
	\end{pmatrix}}.
\end{equation}
For the factors, $\phi_k = 0$,
\begin{equation}
R_\omega(q=1,m>0)
= \begin{pmatrix}
	\frac{1-e^{i(m+1)\omega(\tau+t_\pi)}}{1-e^{i\omega(\tau+t_\pi)}} & 0 & 0 \\
	0 & \frac{1+(-1)^me^{i(m+1)\omega(\tau+t_\pi)}}{1+e^{i\omega(\tau+t_\pi)}} & 0 \\
	0 & 0 & \frac{1+(-1)^me^{i(m+1)\omega(\tau+t_\pi)}}{1+e^{i\omega(\tau+t_\pi)}} \\
\end{pmatrix}R_\omega(m=0).
\end{equation}
Meanwhile, the $q=4$ ghost factor gives
\begin{equation}
\begin{gathered}
R_\omega(q=4,m>0) =  R_\omega(m=0) + \sum_{k=1}^{m} e^{i\omega k(\tau+t_\pi)} (A_k+B_k),\\
\begin{aligned}
A_{k_{\text{even}}} &= \begin{pmatrix}
	e^{i\omega\pqty{\frac{\tau}{2}+t_\pi}} & 1-e^{i\omega\frac{\tau}{2}} & 0\\
	1-e^{i\omega\frac{\tau}{2}} & e^{i\omega\pqty{\frac{\tau}{2}+t_\pi}} & 0\\
	0 & 0 & -\pqty{1-e^{i\omega\frac{\tau}{2}}\pqty{1+e^{i\omega t_\pi}}}
\end{pmatrix} \\
A_{k_{\text{odd}}} &= \begin{pmatrix}
	1-e^{i\omega\frac{\tau}{2}} & e^{i\omega\pqty{\frac{\tau}{2}+t_\pi}} & 0\\
	e^{i\omega\pqty{\frac{\tau}{2}+t_\pi}} & 1-e^{i\omega\frac{\tau}{2}} & 0\\
	0 & 0 & 1-e^{i\omega\frac{\tau}{2}}\pqty{1+e^{i\omega t_\pi}}
\end{pmatrix}
\end{aligned}\\
B_k ={\scriptstyle
e^{i\omega\frac{\tau}{2}}\frac{1+e^{i\omega t_\pi}}{\omega^2-\Omega^2}\begin{pmatrix}
-\frac{1}{2}\pqty{\frac{\omega^2-\Omega^2}{1+e^{i\omega t_\pi}}\pqty{1-e^{i\omega t_\pi}} - (-1)^k \omega^2} &
	-\frac{1}{2}\pqty{\frac{\omega^2-\Omega^2}{1+e^{i\omega t_\pi}}\pqty{1-e^{i\omega t_\pi}} + (-1)^k \omega^2} &
	\frac{q_k}{\sqrt{2}} i\omega\Omega\\
-\frac{1}{2}\pqty{\frac{\omega^2-\Omega^2}{1+e^{i\omega t_\pi}}\pqty{1-e^{i\omega t_\pi}} + (-1)^k \omega^2} &
	-\frac{1}{2}\pqty{\frac{\omega^2-\Omega^2}{1+e^{i\omega t_\pi}}\pqty{1-e^{i\omega t_\pi}} - (-1)^k \omega^2} &
	-\frac{q_k}{\sqrt{2}} i\omega\Omega\\
-\frac{q_k}{\sqrt{2}} i\omega\Omega &
	\frac{q_k}{\sqrt{2}} i\omega\Omega &
	-\omega^2
\end{pmatrix}}\\
\text{where } q_k = \begin{cases}
	1 & \text{if }(k-1)\bmod 4 = 0,1\\
	-1 & \text{if }(k-1)\bmod 4 = 2,3
\end{cases}
\end{gathered}
\end{equation}
\end{widetext}
While the analytical expressions for $g_{ij}(\omega;m) = \frac{1}{\omega^2}\left[R_\omega(\omega;m)R^\dagger_\omega(\omega;m)\right]_{ij}$ become cumbersome to work out, they can be calculated numerically with ease by using the above formulas. In practice, the noise we measure will be data which might not follow the spectrum of a simple distribution, so the integration will be done numerically in any case.

\end{document}